# POWER SYSTEM RESOURCE EXPANSION PLANNING

*A dissertation submitted in partial fulfillment of the requirement for the award of the Degree of*

**MASTER OF TECHNOLOGY**
*in*
**POWER SYSTEMS**

*by*

**SOHOM  DATTA**
**2009EES3013**

*Under the Guidance of*

**PROF. P. R. BIJWE**

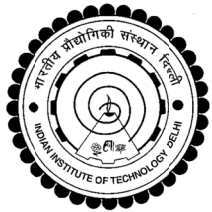

**DEPARTMENT OF ELECTRICAL ENGINEERING**

**INDIAN INSTITUTE OF TECHNOLOGY DELHI**

**MAY 2011**

# CONTENTS

iv

v

# LIST OF FIGURES

# LIST OF TABLES

# CHAPTER 1

## INTRODUCTION

**1.1    GENERAL**

Power System Resource Planning is the recurrent process of studying and determining what facilities and procedures should be provided to satisfy and promote appropriate future demands for electricity. The electric power system as planned should meet or balance societal goals. These include availability of electricity to all potential users at lowest possible cost, minimum environmental damage, high levels of safety and reliability, etc. Plans should be technically and financially feasible. Plans also should achieve the objectives the entity doing the planning, including minimizing risk.

It is well known that most of the optimization problems in power system planning are complex, large scale, hard, nonlinear combinatorial problems of mixed integer nature. This means the number of solutions to be evaluated grows exponentially with the system size, that is, a non-polynomial type (NP-complete) with a large number of local optimal solutions, which makes the solution space a potentially high multimodal landscape.

The very nature of the planning exercise has led to the subdivision of the problem into sub-problems that have gained autonomy of their own. For instance the following problems are referred:

(a) Long-term Generation Expansion Planning, where a capacity addition plan must be produced.

(b) Network expansion plan, either at transmission or at distribution level, typically adopting a DC model to evaluate power flows, where a plan for network additions and reinforcements must be produced.

(c) Reactive power planning, where an investment plan for new reactive sources at selected load buses must be produced, which nowadays is

useful both at transmission and distribution level, due to growing importance of distributed generation.

Generation Expansion Planning (GEP) is one of the most important decision-making activities in electric utilities. In classical vertically integrated power systems, this was a function of electric utilities, but in unbundled organized systems, GEP is still important as a means of developing reference plans for system growth, even if the construction of new power plants depends on the initiative of private investors.

The GEP problem is a highly constrained nonlinear discrete dynamic optimization problem that can only be fully solved by complete enumeration in its nature. Therefore, every possible combination of candidate options over a planning horizon must be examined to get the optimal plan, which is computationally impossible in a real-world GEP problem.

The objective of Transmission Network Expansion Planning (TNEP) consists of determining an optimal expansion plan of the electric network, with circuits that satisfy the operational conditions for a forecasted demand growth under a particular generation expansion plan (GEP).

Network expansions, both at distribution and transmission level, may be studied by static or dynamic models. A static model tries to discover an optimal network structure (where and which type of new equipment should be installed in an optimal way that minimizes the installation and operation costs) for a given scenario of generation and load in a long-term context. A dynamic model is more complex and it aims at, besides answering the question of where and which, defining when to install the network additions, creating therefore a plan of investment along successive periods of time.

This is a large scale, non-linear, mixed integer, non-convex optimisation problem. The problem is very complex and computationally demanding because of large number of options to be investigated and the discrete nature of the optimisation variables. Further, the number of options to be analysed increases exponentially with the size of the system, hence the problem is non-polynomial time-hard. Conventional optimisation techniques provide very successful strategies to obtain global optimum in simple and ideal models. However, real-world engineering optimisation problems like TNEP are very complex and difficult to solve with these methods. In view of this, many new meta-heuristic techniques have been proposed for TNEP in last few years because of their ability to find global optimal solutions for such combinatorial problem.

Reactive power planning (RPP) or VAr planning involves determination of optimal location and determination of types and sizes of the new reactive power sources to satisfy the voltage constraints during normal and contingency states. The objectives may include many cost functions such as variable reactive power source cost, fixed reactive power source cost, real power losses, and fuel cost. The deviation from a given voltage schedule, voltage stability margin, or even a combination of different objectives may also be considered.

Traditionally, the locations for placing new VAr sources were either simply estimated or directly assumed. Recent research has presented some rigorous optimization-based methods to address RPP. Due to the complicated objective functions, constraints, and solution algorithms, RPP is identified as one of the most challenging problems in power systems. It is large scale mixed integer nonlinear problem.

The constraints for this RPP problem are even more complicated than the objective functions. Conventional constraints may include the normal state (base case) power-flow limits and the contingency state power flow limits.

The mathematical solution of RPP is also very challenging due to a large number of variables and uncertain parameters. Based on the objective and constraint formulation, RPP could be mathematically formulated with variables and equations that are partially discrete, partially continuous, non-differentiable, and nonlinear. There are no known ways to find the exact global solution for this complicated optimization problem in a reasonable time.

The integer variables appear in the formulation with the mathematical representation of

- The installation or fixed cost of new reactive sources at the different locations,
- The discrete availability of sizes or capacities of the reactive sources, and
- The discrete characteristics of the transformer tap positions.

Reactive power and voltage control plays an important role in supporting the real power transfer across a large scale transmission system. In open access system the importance of this support is even greater as the power transfer is increased and the associated voltages then become a bottle neck in preventing additional power transfer. In simple terms the most important aim of reactive power planning is to determine the sufficient amount of correct location of reactive power support in order to maintain secure voltage profile. Reactive power support is generally provided by switching of shunt reactors the positioning of transformer taps and reactive power output of generators.

For all these problems, early attempts to solve them by applying mathematical analytical models had only a limited success and in some cases with no real-world application. In practice, the use of heuristics allowed a more realistic representation of the system characteristics, constraints, and objectives of the decision makers, even if potentially sacrificing optimality.

The emergence of meta-heuristics has given robustness to the non-analytical methods, because of the rationale behind them. Besides, evolutionary algorithms have provided a higher degree of confidence in a stochastic convergence to optimum and have supported this confidence with a mathematical background explaining not only how they achieve convergence but also how to improve the convergence rate.

Recently, some authors have proposed to solve these discrete variable optimization problems using Interior Point Methods. This technique requires conversion of continuous variables to discrete variables using sigmoid functions.

This chapter is organized as follows: The various important contributions made by different authors in the study of power system expansion planning have been discussed in section (1.2). A brief motivation behind this thesis is presented in section (1.3). Section (1.4) presents the structural outline of this report.

## 1.2    LITERATURE SURVEY

To solve the Generation Expansion Planning problem, a number of salient methods have been successfully applied during the past decades. Bloom[5] applied a mathematical programming technique using a decomposition method, and solved it in a continuous space. Park *et al.*[6]  applied the Pontryagin's maximum principle whose solution also lies in a continuous space. Although the above-mentioned mathematical programming methods have their own advantages, they possess one or both of the following drawbacks in solving a GEP problem. That is, they treat decision variables in a continuous space and there is no guarantee to get the global optimum since the problem is not mathematically convex.

Dynamic programming based framework is one of the most widely used algorithms in GEP. However, so called "the curse of dimensionality" has interrupted direct application of the conventional full DP in practical GEP problems. David and Zhao[7] developed a heuristic-based DP and applied the fuzzy set theory to reduce the number of states.

Recently, Fukuyama and Chiang[8] applied genetic algorithm (GA) to solve sample GEP problems, and showed promising results. Park *et al.*[1] have used improved version of GA to solve GEP as simple GA has problems like premature convergence. Kannan, Slochanal and Padhy[2] have compared the different metaheuristic techniques like Genetic Algorithm, Ant Colony Optimization, Particle Swarm Optimization, Tabu Search, Simulated Annealing and Hybrid Approach to solve the GEP problem. Kannan, Baskar, McCalley and Murugan[4] have solved the GEP using an advanced version of GA called nondominated sorting genetic algorithm version II (NSGA-II).

A methodology for calculating the optimal generation capacity reserve and siting in power system planning considering transmission network is proposed in [2]. A model and algorithms to assess the performance of a certain plan of composite generation and transmission system expansion are presented in [3]. The transmission constraints have been included into integrated resource planning with static model in [4]. Anders proposed the generation expansion planning (GEP) model that includes the transmission lines with reliability constraints, modeled both in discrete and continuous case and solved using LP [5]. The GEP with power flow constraints has been solved using evolutionary strategy approach [6].

GA is a search algorithm based on the hypothesis of natural selections and natural genetics. Recently, a global optimization technique using GA has been successfully applied to various areas of power system such as economic dispatch [9], [10], unit commitment [11], [12], reactive power planning [13]–[15], and power plant control [16]. GA-based approaches for least-cost GEP have several advantages. Naturally, they can not only treat the discrete variables but also overcome the dimensionality problem. In addition, they have the capability to search for the global optimum or quasioptimums within a reasonable computation time.

Transmission Network Planning using Linear programming was presented by Garver [35], where the use of linear programming for network analysis to determine where capacity shortages exist and, most importantly, where to add new circuits to relieve the shortages is presented.

Test systems and mathematical models for transmission network expansion planning are provided by Romero et.al in [34], where the main mathematical formulations used in transmission expansion studies-transportation models, hybrid models, DC power flow models, and disjunctive models are also summarized and compared. The main algorithm families are reviewed viz. analytical, combinatorial and heuristic approaches.

A simulated annealing (SA) approach for TNEP is proposed in [29] for long-term TNEP. A parallel Tabu search algorithm for TNEP is discussed in [30]. An improved GA is proposed for TNEP in [31]. A GA based approach for multistage and coordinated planning of transmission expansions is presented in [32]. A specialized GA is proposed in [33] for static and multistage TNEP.

In literatures [22–27], the models and algorithms to integrate transmission line expansion into GEP have been developed. [25] presents a minimum cost assessment method for composite system expansion planning, which can be used to consider generation expansion and transmission expansion simultaneously. The minimization model proposed in [25] to incorporate both operating and outage costs can recognize different customer damage functions at different load buses and includes the duration of the simulated contingency system states. [17] uses network topology methods for optimization of the composite generation expansion planning. [18] addresses the problem of a multiyear security constrained hybrid generation-transmission expansion planning assuming the overall generation requirements of a network are known along the planning horizon, but their allocations are unknown. By allocating the overall generation capacity among the grid nodes, and determining the new transmission element additions along the planning horizon, the overall cost of the system is minimized.

AC model based TNEP is presented by Rider, Garcia & Romero in [36] where constructive heuristic algorithm aimed at obtaining an excellent quality solution for the problem is given.

The majority of the RPP objectives were to provide the least cost of new reactive power supplies. Many variants of this objective include the cost of real power losses or the fuel cost. In addition, some technical indices such as deviation from a given voltage schedule or the security margin may be used as objectives for optimization. The detailed discussions are presented as follows.

Generally, there are two VAr source cost models for minimization. The first formulation is to model VAr source costs which represent a linear function with no fixed cost. Apparently, this model considers only the variable cost relevant to the rating of the newly installed VAr source and ignores the fixed installation cost. This formulation would always bias a solution toward placement of several smaller sizes sources instead of a small number of larger ones. A better formulation with fixed and variable costs [38]–[40] is to consider the fixed cost, ($/hour), which is the lifetime fixed cost prorated to per hour, in addition to the incremental/variable cost. This is a more realistic model of VAr cost, but this would complicate the problem from a nonlinear programming (NLP) to a Mixed-integer NLP (MINLP), because there is a binary variable to indicate whether the VAr source will be actually installed or not. The slight difference in the cost model, however, leads to dramatic difference in the optimization model and the corresponding mathematics technique to solve it. As a result, the RPP model with the first VAr cost function as an objective is a traditional LP or NLP problem. However, the second one is a MINLP problem, and some special techniques are needed for it. The objective under this category may be divided into two groups. To minimize variable cost and Power loss [41], [42], [49], [51], [52] or to minimize complete cost function and power loss.[43]-[46], [50].

The algorithms for solving reactive power planning may be classified into three groups: Conventional optimization methods including nonlinear programming (LP), Mixed integer programming, Intelligent searches such as simulated annealing, evolutionary algorithms (EAs), and tabu search and fuzzy set applications to address uncertainties in objectives and constraints. Heuristic methods such as Simulated annealing [47], [48], Evolutionary algorithms and Tabu search are capable of finding global optimum even though they are computationally inefficient.

In the existing TNEP approaches, the transmission network planning problem is first handled using both classical optimization techniques [35] and evolutionary algorithms [29–33]. Subsequently, the expanded network is reinforced considering the reactive power source allocation (RPP) [38–52]. It should be noted that there are a few reports in existing planning methods discussing both stages in an integrated way.

A combinatorial mathematical model in tandem with a metaheuristic technique for solving transmission network expansion planning (TNEP) using an AC model associated with reactive power planning (RPP) is presented Rahmania, Rashidinejada, Carrenoc, Romerob in [37].

Torres and Quintana [53] solved nonlinear programming problems in power engineering like optimal power flow problems using Interior point Method. They also discussed apart from the algorithm several important issues that are critical to their efficient implementation.

Recently, Carneiro et al.[54] developed a model for OPF algorithm that used the primal–dual interior point technique to determine the best investment strategy in the TNEP problem. In the proposed method, the expansion decision (0 or 1) is mitigated by using a sigmoid function, which is incorporated in the OPF problem through the modified dc power flow equations.

## 1.3 MOTIVATION

In the light of literature survey following problems was considered to be worth exploring.

Transmission Constrained Generation Expansion Planning (TC GEP) is to optimize the least cost generation expansion planning (GEP) subjected to network transmission constraints. GEP without network constraints may seem to be more economical apparently but the optimal plan may be infeasible for the available transmission capacity and may require further reinforcements in the transmission circuits for transfer of power to the load ends.

Traditionally, TNEP is done after getting a generation expansion plan. This may lead to additional transmission lines reinforcement which could have been taken care of by modifying the generation expansion plan. Hence, a novel approach would be Composite Generation Expansion and Transmission Network Expansion Planning which may provide better and economical plans although the optimization problem becomes more complex.

TNEP using DC model finds out the network requirements for feasible power flows in the network. Reactive power flows in the lines are usually considered in the operations problem instead of the planning problem. If AC model and fixed power factor at the load end is considered for TNEP then reactive power sources required during operations will be less.

A novel approach to TNEP and RPP may be the decomposed integration of AC TNEP and RPP which shall allow the transmission lines additions and reactive power sources additions to alter at each step until a mutually feasible solution is reached. Hence the resulting plan will be better than traditional method of reactive planning after transmission planning.

Interior Point (IP) based methods for power system optimization problems have given great results. But because of integer variables associated with planning problems IP methods have not been used in planning problems until the recent past. Sigmoid functions are used for converting continuous variables into discrete ones. TNEP was modelled to solve using Newton based IP method. This technique is very recent for the power system planning problems.

## 1.4    OUTLINE OF THESIS

The present work of analyses and implementation can be divided into:

i) Transmission Constrained Generation Expansion Planning (TC GEP),

ii) Composite Generation Expansion and Transmission Network Expansion Planning (GEP TNEP),

iii) Transmission Network Expansion (TNEP) Planning using AC model,

iv) Composite Transmission Network Expansion Planning (TNEP) and Reactive Power Expansion Planning (RPP) and

v) Transmission Network Planning using Interior-Point Method (IP TNEP)

Hereafter, the chapters which appear in the report have been briefly introduced here.

Chapter 1 introduces the power system resource planning problem. The Chapter presents the brief survey of previous work done concerning the planning problems in power systems problem and the motivation behind this thesis.

Chapter 2 presents the Transmission Constrained Generation Expansion Planning (TC-GEP) problem using Genetic Algorithm. Problem Formulation and detailed theoretical background of the problem has been presented here. The test results of implementation of the above stated methods have been presented.

Chapter 3 presents a novel approach of Composite Generation and Transmission Network Expansion Planning problem. Problem Formulation and detailed theoretical background of the method have been discussed. The chapter finally concludes with the test result of implementation of the method using Genetic Algorithm.

Chapter 4 presents the Transmission Network Expansion (TNEP) Planning using AC model problem. Problem Formulation and detailed theoretical background of the method have been discussed. The chapter finally concludes with the test result of implementation of the method using Genetic Algorithm with and without security constraints.

Chapter 5 presents the novel technique for Composite Transmission Network Expansion Planning (TNEP) and Reactive Power Expansion Planning (RPP) problem. Problem Formulation and detailed theoretical background of the method have been discussed. The chapter finally concludes with the test result of implementation and comparison with TNEP and reactive power planning in traditional disintegrated fashion with and without security constraints.

Chapter 6 presents the Transmission Network Planning using Interior-Point Method (IP TNEP). Problem Formulation and detailed theoretical background of the method have been discussed.

Chapter 7 finally concludes with the important observations and deductions made. A list of possible future work that can be taken up is also given.

A list of references and Appendices has also been added. Appendix I give the system specification details about the test system.

# CHAPTER 2

# TRANSMISSION CONSTRAINED GENERATION EXPANSION PLANNING

## 2.1 INTRODUCTION

GENERATION expansion planning (GEP) is one of the most important decision-making activities in electric utilities. Least-cost GEP is to determine the minimum-cost capacity addition plan (i.e., the type and number of candidate plants) that meets forecasted demand within a pre-specified reliability criterion over a planning horizon.

The solution to the Transmission Constrained Generation Expansion Planning (TC-GEP) problem is finding the most economical generation expansion scheme, achieving certain reliability level to meet out the forecast demand which satisfies transmission and other constraints. The TC-GEP problem is a large scale and challenging problem for the decision makers (to decide upon site, capacity, type of fuel, etc.) as there exist a large number of combinations. This chapter is based on the literature [3].

In this work the TC GEP problem is solved using simple genetic algorithm. Unlike normal GEP network topology has to be considered in TC GEP. The generations of the new and existing generators are obtained by performing lossless economic dispatch (ED). Network flow constraints are modeled by lossless DC power flow. Apart from network flow constraints other constraints for this problem are demand constraint, reliability constraint, fuel mix ratio constraint, maximum construction limit constraints and reserve constraints. The reliability criteria used is LOLP which is discussed later in the chapter. The detailed problem formulation is showed in the next section. The model for GEP used is dynamic and discounted value of costs is referred to the current year.

This chapter is organized as follows: The problem formulation of TC-GEP is discussed in section (2.2). The solution methodology is discussed in detail in section (2.3). Systems used for implementation and the results of these approaches are presented in section (2.4). The last section (2.5) concludes with the important observations.

## 2.2    PROBLEM FORMULATION

The TC-GEP problem is equivalent to finding a set of best decision vectors over a planning horizon that minimizes the investment and operating costs with several constraints.

The cost objective is represented by the following expression:

$$\min_{U_1 \dots U_T} C = \sum_{t=1}^{T} [I(U_t) + M(X_t) - S(U_t)] \tag{2.1}$$

where

$$X_t = X_{t-1} + U_t \qquad t = (1,2, \dots T) \tag{2.2}$$

$$I(U_t) = (1+d)^{-2t'} \sum_{i=1}^{N} (CI_i \times U_{t,i}) \tag{2.3}$$

$$S(U_t) = (1+d)^{-2(T+1)} \sum_{i=1}^{N} \left( CI_i \times \delta_i^{2(T-t+1)} \times U_{t,i} \right) \tag{2.4}$$

$$M(X_t) = \sum_{s'=0}^{1} \left( (1+d)^{-(2.5+t'+s')} \times \sum_{i=1}^{N} \left( (X_{t,i} \times FC_i) + MC_i \times EES_{t,i} \right) \right) \tag{2.5}$$

$$t' = 2(t-1) \ and \ T' = 2 \times T - t'$$

$$U_t = \sum_{i=1}^{N} U_{t,i} \tag{2.6}$$

$$X_t = \sum_{i=1}^{N} X_{t,i} \tag{2.7}$$

$C$      total cost \$;

$U_t$      N-dimensional vector of newly introduced units in the stage t (1 stage =2 years);

$U_{t,i}$      the number of newly introduced units of type *i* in stage *t*;

$X_t$      cumulative capacity vector of existing units in stage *t*,(MW);

$X_{t,i}$      cumulative capacity of existing units in of type *i* stage *t*,(MW);

$I(U_t)$      present value of investment cost of the newly introduced unit at the $t$th stage, \$;

$M(X_t)$   present value of investment cost of the newly introduced unit at the $t$th stage, \$;

$s'$      variable used to indicate that the maintenance cost is calculated at the middle of each year;

$S(U_t)$      present value of salvage value of the newly added unit at $t$th interval, \$;

$d$      discount rate;

$CI_i$      capital investment cost of the $i$th unit, \$;

$\delta_i$      salvage factor of the $i$th unit;

$T$      length of the planning horizon (in stages);

$N$      total number of different types of units;

$FC$      fixed operation and maintenance cost of the units, \$/MW;

$MC$      variable operation and maintenance cost of the units (energy), \$/MWh;

$EES$      expected energy served, MWh;

Constraints are taken as follows:

a) ***Reserve Margin****:* The selected units must satisfy the minimum and maximum reserve margin.

$$(1 + R_{min}) \times D_t \leq \sum_{i=1}^{N} X_{t,i} \leq (1 + R_{max}) \times D_t \qquad (2.8)$$

$R_{min}$      minimum reserve margin;

$R_{max}$      maximum reserve margin;

$D_t$      demand at the $t$th stage in MW

b) *Upper Construction Limit*: Let $U_t$ represent the units to be committed in the expansion plan at stage $t$ that must satisfy

$$0 \leq U_t \leq U_{max,t} \tag{2.9}$$

where $U_{max,t}$ is the maximum construction capacity of the units at stage t.

c) *Demand*: The selected units must satisfy the demand

$$\sum_{i=1}^{N} X_{t,i} \geq D_t \tag{2.10}$$

d) *Reliability criteria*: The selected units must along with the existing units must satisfy a reliability criterion on loss of load probability

$$LOLP(X_t) \leq \varepsilon \tag{2.11}$$

where $\varepsilon$ is the reliability criterion for maximum allowable LOLP

e) *Fuel Mix Ratio*: The GEP has different types of generating units such as coal, liquefied natural gas (LNG), oil, and nuclear. The selected units along with the existing units of each type must satisfy the fuel mix ratio

$$M_{min}^{j} \leq {X_{t,j}}\Big/{\sum_{i=1}^{N} X_{t,i}} \leq M_{max}^{j} \tag{2.12}$$

where

$M_{min}^{j}$          minimum fuel mix ratio of $j^{\text{th}}$ type;

$M_{max}^{j}$          maximum fuel mix ratio of $j^{\text{th}}$ type;

$j$          type of the unit (e.g., oil, LNG, coal, nuclear).

f) *Power Flow Constraints:* Lossless Economic Dispatch is performed and DC load flow is performed to check if all the power flow constraints are satisfied

*Active Power Generation Limit:* $P_{Gi}{}^{\min} \leq P_{Gi} \leq P_{Gi}{}^{\max}$ $\tag{2.13}$

*Active Power Balance Equation:* $\sum P_{Gi} - P_D = 0$ $\tag{2.14}$

*Line Power Flow Limits*: $fl_{ij} \leq fl_{ij}{}^{\max}$ $\tag{2.15}$

## 2.3     THEORETICAL BACKGROUND

The objective function is the tripartite discounted costs over a planning horizon. It is composed of discounted investment costs, expected fuel and O&M costs, salvage value. LOLP evaluation and basics of Genetic Algorithm are discussed in the following articles.

### 2.3.1     EVALUATION OF LOLP

Let S be a random variable representing the available power supply, and let L be the daily load peak. If the peak load is regarded as known, then

$$LOLP = \Pr(S < L) \tag{2.15}$$

Usually the system capacity is designed to have a very small LOLP.

To estimate the LOLP, the first step is the estimation of S.

We assume that any station has two possible working states: operational or down. The probability of the station being down is known as the Forced Outage Rate (FOR). The FOR can be estimated from the station's history. The authors have estimated the FOR for all the fuel stations, using records from the electric companies that contain the exact minute of the beginning and end of every failure in the last five years. There are also theoretical FORs which are obtained from the plant's designers on the basis of technical characteristics of the plants. The estimated FORs are in general smaller than the theoretical ones, so that the reliability of the system is underestimated for security purposes.

Let $X_i$ be a random variable representing the available power that can be supplied by the $i^{th}$ plant.

$$X_i = \begin{cases} 0, & Pr(X_i = 0) = FOR_i \\ C_i, & Pr(X_i = C_i) = 1 - FOR_i \end{cases} \tag{2.16}$$

where $C_i$ is the capacity and $FOR_i$ the Forced Outage Rate of plant i.

Now the available supply $S = \sum_{i=1}^{n} X_i$ \hfill (2.17)

is the sum of n independent random variables. Considering k variables, if $G_k(x)$ is the distribution function of $Z_k = \sum_{i=1}^{k} X_i$ \hfill (2.18)

Then

$G_{k+1}(x) = Pr(Z_{k+1} \leq x) =$

$Pr(Z_k \leq x \, | X_{k+1} = 0) \, Pr(X_{k+1} = 0)$

$\qquad\qquad + Pr(Z_k \leq x - C_{k+1} | X_{k+1} = C_{k+1}) \, Pr(X_{k+1} = C_{k+1})$

$= G_k(x)FOR_{k+1} + G_k(x - C_{k+1})(1 - FOR_{k+1})$  (2.19)

### 2.3.2  GENETIC ALGORITHM

Genetic algorithm is a meta-heuristic procedure which starts from a random initial guess and gradually converges to the optimal solution. It operates on a string structures called chromosomes, typically a concentrated list of binary digits representing an encoding of the control parameters of a given problem. The encoding scheme of chromosome is made of a binary string which contains a set of active and inactive alleles (genes).

For better understanding, the encoding and decoding scheme of GA can be explained as below.

**Encoding:** This is the method by which variables are transformed into binary strings. Bit length of the strings depends on the resolution of the variable required for solving the particular optimization problem.

**Decoding:** GA's are not restricted to use integer values only, they can be assigned any non integer values also by choosing the upper and lower bounds and the string length.

The decoded value of a variable can be given as

$$x_i = x_i^{\min} + \frac{x_i^{\max} - x_i^{\min}}{2^{l_i} - 1} DV(s_i)$$  (2.20)

$l_i$ is the string length (in above example it is 4) used to code the $i^{th}$ variable, and $DV(s_i)$ is the decoded value of the string $s_i$. $x_i^{\min}$ - $x_i^{\max}$ : allowable range of $i^{th}$ variable.

GA consists of three basic operators, known as 1) reproduction 2) crossover and 3) mutation which are discussed in brief below.

**Reproduction:** It comprises of forming a new population, usually with the same total number of chromosomes, by selecting from members of the current population following a particular scheme. The higher the fitness, more are the chances of its selection for the next generation.

Hence, the purpose of reproduction is

1) to identify the good solutions in a population.

2) Make multiple copies of good solutions.

3) Eliminate bad solutions so that multiple copies of good solutions can be placed in the population.

There are many strategies for selection e.g. tournament selection, proportionate selection and ranking selection.

*Tournament selection*: In tournament selection, tournaments are played between the two solutions and the better solution is chosen and placed in the mating pool. Two other solutions are picked again and another slot in the mating pool is filled with better solutions. Each solution will be made to participate in exactly two tournaments. The best solution in a population will win both the times, thereby making two copies of it in the new population and the worst solution will lose in both tournaments and will be eliminated from the population.

*Proportionate selection:* In this selection operator the solutions are assigned copies, the number of which is proportional to their fitness value. If average fitness of all population members is $f_{avg}$, a solution with a fitness $fi$ gets an expected value of $f_i/f_{avg}$ number of copies. The implementation of this type of procedure can be thought of as a roulette-wheel mechanism.

*Ranking selection:* In ranking selection, the solutions are ranked according to their fitness and then the multiple copies of good solutions are found.

**Crossover:** A crossover operator is applied next to the strings of the mating pool. The crossover operator is mainly responsible for the global search property of the GA. Reproduction operator only makes copies of good solutions and eliminates the bad solutions. The creation of new solutions is mainly done by the crossover and mutation operators. Two strings are picked at random and some portions of the strings are interchanged between the strings to create two new strings. Crossover can be a single point crossover or multipoint crossover.

**Mutation**: The bitwise mutation operator changes 0 to 1 and vice versa, with a mutation probability of $p_m$. The need for mutation is to keep diversity in the population.

**Elitism:** Strings of a population with high fitness are copied to the new population otherwise there may be loss of good solutions. Only a few strings are kept as elites.

## 2.4 SOLUTION METHODOLOGY

Simple Genetic Algorithm is used for solving the transmission constrained generation expansion planning problem.

**String Selection:**

| Stage 1 | Stage 2 | ... | ... | Stage t | ... | ... | Stage T-1 | Stage T |
|---------|---------|-----|-----|---------|-----|-----|-----------|---------|

**Sub-string** (Addition of i[th] candidate at each stage t)

| $U_{t,1}$ | $U_{t,2}$ | $U_{t,3}$ | $U_{t,4}$ | $U_{t,5}$ |
|-----------|-----------|-----------|-----------|-----------|

- String length for each unit i at each stage t is taken as 2 bits.
- Therefore, at each stage for 21 different units each substring length becomes 2×21 = 42
- For 3 stages or 6 year planning horizon string length becomes 3×42 = 126

**Fitness Function:**

$$fitness = \frac{\alpha}{1+J} \qquad (2.21)$$

Where α is a constant, J is the objective function and penalties for any constraint violation

**Parameters for GEP:**

d = discount rate = 8.5%
LOLP criteria = 0.01
Min. Reserve Margin = 20% Max. Reserve Margin = 60%
Parent Selection Method: Tournament Selection
Crossover Technique: Uniform Crossover

**Table 2.1 Genetic Algorithm Parameters**

| GA Parameters | Value |
|---------------|-------|
| **Population Size** | 100 |
| **Maximum Generations / Iterations** | 1000 |
| **Probabilities of Crossover and Mutation** | 0.9, 0.01 |
| **Number of Elite Strings** | 5 |

## 2.5    RESULTS

This section presents the results obtained after the implementation of the TC GEP problem using Genetic Algorithm. The modified IEEE 24 bus Reliability Test System has been used for implementation of the problem discussed in the previous section. DC load flow is used for finding network constraints. Lossless Economic Dispatch is performed for finding the generation schedules using Matlab FMINCON. Monte Carlo Simulations are performed with 400000 samples for validation purposes.

### 2.5.1   IEEE 24 BUS SYSTEM

**Figure 2.1** and **Figure 2.2** shows the maximization of fitness function and minimization of objective respectively. Technical and economic data are provided in **Table-2.2-2.4**. Results are tabulated in **Table 2.5-2.9**.

**Table 2.2 Technical and Economic Data of the Existing Plants**

| Name (Fuel Type) | No. of Units | Unit Capacity (MW) | FOR(%) | Operating Cost ($/KWh) | Fixed Cost (S/KW-Month) |
|---|---|---|---|---|---|
| LNG #1 | 1 | 591 | 7 | 0.024 | 2.25 |
| Oil #1 | 1 | 40 | 3 | 0.043 | 4.52 |
| Oil #2 | 1 | 40 | 10 | 0.038 | 1.63 |
| Oil #3 | 1 | 300 | 10 | 0.04 | 1.63 |
| Coal #1 | 1 | 152 | 15 | 0.023 | 6.65 |
| Coal #2 | 1 | 152 | 9 | 0.019 | 2.81 |
| Coal #3 | 1 | 155 | 8.5 | 0.015 | 2.81 |
| Coal #4 | 1 | 155 | 8.5 | 0.015 | 2.81 |
| Coal #5 | 1 | 300 | 8.5 | 0.015 | 2.81 |
| Coal #6 | 1 | 660 | 8.5 | 0.015 | 2.81 |
| Nuclear #1 | 1 | 400 | 9 | 0.005 | 4.94 |
| Nuclear #2 | 1 | 400 | 8.8 | 0.005 | 4.63 |

**Table 2.3 Forecasted Peak Demand**

| Stage | 0 | 1 | 2 | 3 |
|---|---|---|---|---|
| Year | 2007 | 2009 | 2011 | 2013 |
| Peak(MW) | 3238 | 3885.6 | 4662.7 | 5595.3 |

**Table 2.4 Technical and Economic Data of the Candidate Plants**

| Candidate Type | Construction Upper Limit | Unit Capacity (MW) | FOR(%) | Operating Cost ($/KWh) | Fixed Cost (S/KW-Month) | Capital Cost ($/KW) | Life Time (Yrs) |
|---|---|---|---|---|---|---|---|
| LNG #1 | 5 | 100 | 7 | 0.021 | 2.2 | 812.5 | 25 |
| LNG #2 | 3 | 100 | 7 | 0.021 | 2.2 | 812.5 | 25 |
| LNG #3 | 2 | 50 | 7 | 0.021 | 2.2 | 812.5 | 25 |
| LNG #4 | 5 | 200 | 7 | 0.021 | 2.2 | 812.5 | 25 |
| LNG #5 | 6 | 200 | 7 | 0.021 | 2.2 | 812.5 | 25 |
| LNG #6 | 6 | 200 | 7 | 0.021 | 2.2 | 812.5 | 25 |
| Oil #1 | 3 | 100 | 10 | 0.035 | 0.9 | 500 | 20 |
| Oil #2 | 5 | 100 | 10 | 0.035 | 0.9 | 500 | 20 |
| Oil #3 | 4 | 50 | 10 | 0.035 | 0.9 | 500 | 20 |
| Oil #4 | 3 | 100 | 10 | 0.035 | 0.9 | 500 | 20 |
| Coal #1 | 5 | 100 | 9.5 | 0.014 | 2.75 | 1062.5 | 25 |
| Coal #2 | 5 | 100 | 9.5 | 0.014 | 2.75 | 1062.5 | 25 |
| Coal #3 | 5 | 100 | 9.5 | 0.014 | 2.75 | 1062.5 | 25 |
| Coal #4 | 6 | 200 | 9.5 | 0.014 | 2.75 | 1062.5 | 25 |
| Coal #5 | 6 | 200 | 9.5 | 0.014 | 2.75 | 1062.5 | 25 |
| Coal #6 | 5 | 100 | 9.5 | 0.014 | 2.75 | 1062.5 | 25 |
| Coal #7 | 4 | 200 | 9.5 | 0.014 | 2.75 | 1062.5 | 25 |
| Nuclear #1 | 3 | 100 | 9 | 0.004 | 4.6 | 1625 | 25 |
| Nuclear #2 | 5 | 100 | 9 | 0.004 | 4.6 | 1625 | 25 |
| Nuclear #3 | 5 | 100 | 9 | 0.004 | 4.6 | 1625 | 25 |
| Nuclear #4 | 5 | 100 | 9 | 0.004 | 4.6 | 1625 | 25 |

### 2.5.1 Results for 6 Years Planning Horizon (3 stages) TC GEP

| *6 Years Planning Horizon (3 stages) TC GEP* |
| :---: |
| **Minimized Total Cost : 3.0821×10⁹ \$** |

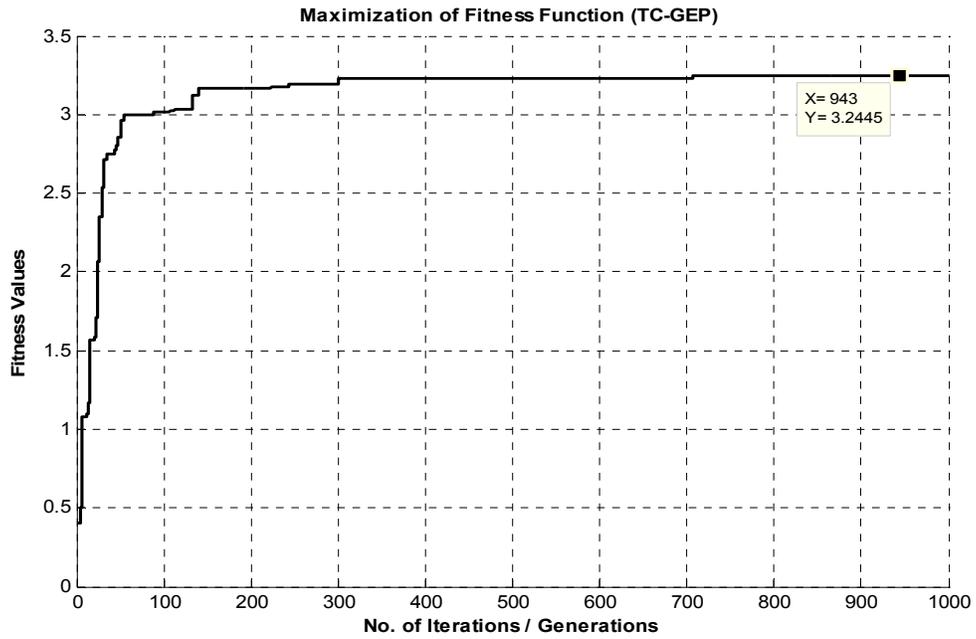

**Fig 2.1 Maximization of Fitness Function (TC GEP)**

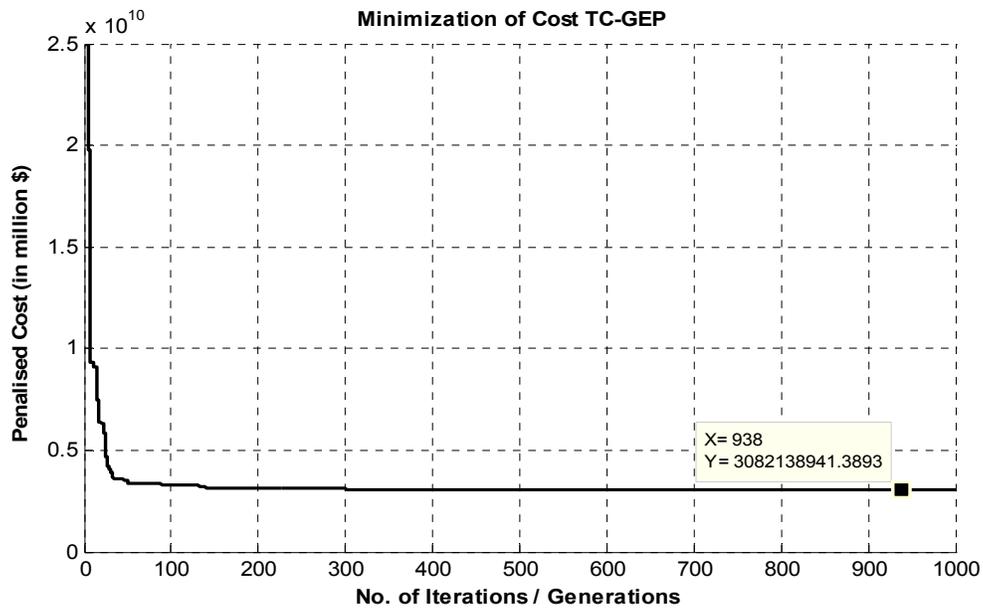

**Fig 2.2 Minimization of Objective Function (TC GEP)**

**Table 2.5** Stage-wise capacity addition of new units of candidate plants

| SL. No. | Stage 1 | Stage 2 | Stage 3 | Total |
|---------|---------|---------|---------|-------|
| LNG #1 | 0 | 0 | 0 | 0 |
| LNG #2 | 1 | 0 | 1 | 2 |
| LNG #3 | 0 | 1 | 0 | 1 |
| LNG #4 | 0 | 0 | 2 | 2 |
| LNG #5 | 3 | 0 | 0 | 3 |
| LNG #6 | 0 | 0 | 0 | 0 |
| Oil #1 | 1 | 1 | 1 | 3 |
| Oil #2 | 2 | 0 | 0 | 2 |
| Oil #3 | 1 | 2 | 0 | 3 |
| Oil #4 | 2 | 1 | 0 | 3 |
| Coal #1 | 2 | 0 | 2 | 4 |
| Coal #2 | 0 | 0 | 0 | 0 |
| Coal #3 | 0 | 0 | 0 | 0 |
| Coal #4 | 0 | 0 | 0 | 0 |
| Coal #5 | 0 | 0 | 0 | 0 |
| Coal #6 | 0 | 0 | 0 | 0 |
| Coal #7 | 1 | 1 | 0 | 2 |
| Nuclear #1 | 0 | 0 | 1 | 1 |
| Nuclear #2 | 0 | 0 | 0 | 0 |
| Nuclear #3 | 0 | 0 | 0 | 0 |
| Nuclear #4 | 0 | 0 | 0 | 0 |

**Table 2.6 Generation and Reserves at Each Stage**

| Stage | Peak MW Demand at each stage | Total Generation Capacity at each stage (MW) | Reserve at each stage (MW) |
|-------|------------------------------|----------------------------------------------|----------------------------|
| 0 | 3238 | 3345 | 107 |
| 1 | 3885.6 | 4995 | 1109.4 |
| 2 | 4662.7 | 6445 | 1782.3 |
| 3 | 5595.3 | 8145 | 2549.7 |

### 2.5.1 Results for 6 Year Planning Horizon (3 stages) simple GEP without transmission constraints

| 6 Year Planning Horizon (3 stages) simple GEP without transmission constraints |
| --- |
| Minimized Total Cost: 2.8775×10⁹ $ |

**Table 2.8 Stage-wise capacity addition of new units of candidate plants**

| SL. No. | Stage 1 | Stage 2 | Stage 3 | Total |
| --- | --- | --- | --- | --- |
| LNG #1 | 0 | 0 | 2 | 2 |
| LNG #2 | 1 | 0 | 1 | 2 |
| LNG #3 | 0 | 0 | 1 | 1 |
| LNG #4 | 2 | 0 | 0 | 2 |
| LNG #5 | 2 | 0 | 2 | 4 |
| LNG #6 | 0 | 0 | 0 | 0 |
| Oil #1 | 2 | 1 | 0 | 3 |
| Oil #2 | 2 | 0 | 0 | 2 |
| Oil #3 | 1 | 0 | 0 | 1 |
| Oil #4 | 1 | 1 | 1 | 3 |
| Coal #1 | 0 | 0 | 0 | 0 |
| Coal #2 | 0 | 0 | 0 | 0 |
| Coal #3 | 0 | 0 | 0 | 0 |
| Coal #4 | 0 | 0 | 0 | 0 |
| Coal #5 | 0 | 0 | 0 | 0 |
| Coal #6 | 0 | 0 | 0 | 0 |
| Coal #7 | 1 | 1 | 1 | 3 |
| Nuclear #1 | 0 | 0 | 0 | 0 |
| Nuclear #2 | 0 | 0 | 0 | 0 |
| Nuclear #3 | 0 | 0 | 0 | 0 |
| Nuclear #4 | 0 | 0 | 0 | 0 |

**Table 2.9 Generation and Reserves at Each Stage**

| Stage | Peak MW Demand at each stage | Total Generation Capacity at each stage (MW) | Reserve at each stage (MW) |
| --- | --- | --- | --- |
| 0 | 3238 | 3345 | 107 |
| 1 | 3885.6 | 4945 | 1059.4 |
| 2 | 4662.7 | 5545 | 882.3 |
| 3 | 5595.3 | 6595 | 999.7 |

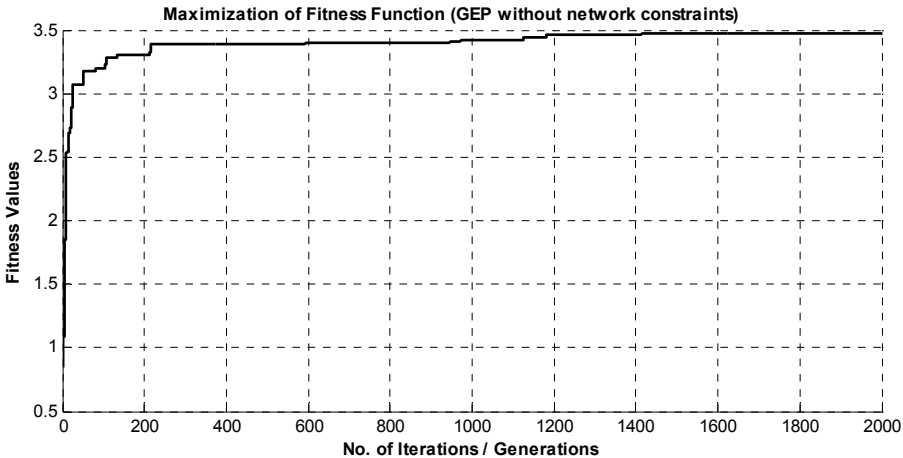

**Fig 2.3 Maximization of Fitness Function (Unconstrained GEP)**

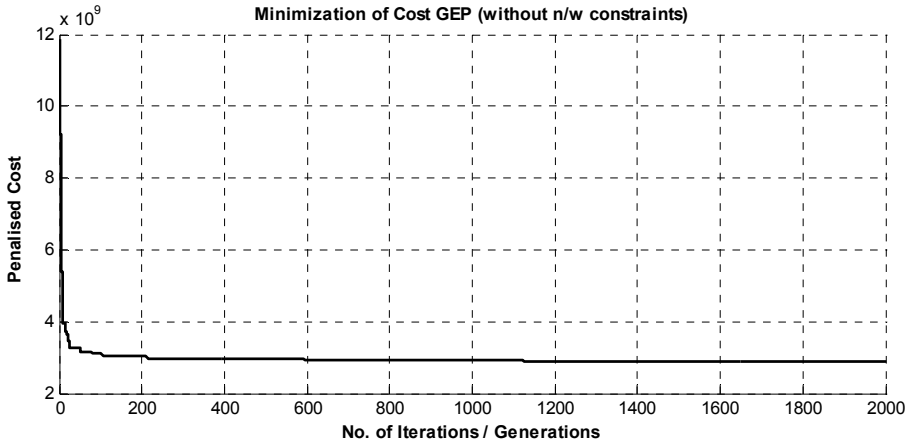

**Fig 2.4 Minimization of Objective Function (Unconstrained GEP)**

**Table 2.7 Line Flows in the System after planning using TC GEP and simple GEP**

| From Bus | To Bus | Line Flows (pu) TC GEP | Line Flows (pu) GEP w/o line constraints | Line Flow Limits (pu) |
|---|---|---|---|---|
| 1 | 2 | -0.1355 | -0.15527 | 0.175 |
| 1 | 3 | 0.1024 | 0.101565 | 0.175 |
| **1** | **5** | **0.1802** | **0.200799** | **0.2** |
| 2 | 4 | 0.157 | 0.178167 | 0.4 |
| 2 | 6 | 0.0583 | 0.117356 | 0.2 |
| 3 | 9 | -0.0216 | 0.018568 | 0.5 |
| 3 | 24 | -0.2303 | -0.27124 | 0.4 |
| 4 | 9 | 0.0101 | 0.031287 | 0.5 |
| 5 | 10 | 0.0402 | 0.060831 | 0.4 |
| 6 | 10 | 0.1564 | 0.032473 | 0.4 |
| 7 | 8 | 0.2605 | 0.173376 | 0.4 |
| 7 | 9 | 0.16 | -0.0358 | 0.5 |
| 8 | 9 | -0.0374 | -0.07779 | 0.5 |
| 8 | 10 | -0.0391 | -0.08579 | 0.3 |
| 9 | 11 | -0.0965 | -0.18206 | 0.5 |
| **9** | **12** | **-0.1345** | **-0.22381** | **0.2** |
| 10 | 11 | -0.0931 | -0.16632 | 0.4 |
| 10 | 12 | -0.1312 | -0.20806 | 0.5 |
| 11 | 13 | -0.1716 | -0.22699 | 0.5 |
| 11 | 14 | -0.018 | -0.12139 | 0.2 |
| 12 | 13 | -0.1046 | -0.15341 | 0.2 |
| 12 | 23 | -0.1611 | -0.27847 | 0.4 |
| 13 | 23 | -0.1224 | -0.22656 | 0.4 |
| **14** | **16** | **-0.3981** | **-0.50155** | **0.4** |
| 15 | 16 | 0.2528 | -0.04714 | 0.4 |
| 15 | 21 | -0.1921 | -0.30411 | 0.4 |
| 15 | 21 | -0.1921 | -0.30411 | 0.4 |
| 15 | 24 | 0.2303 | 0.271243 | 0.5 |

| 16 | 17 | -0.3115 | -0.37042 | 0.5 |
| 16 | 19 | 0.2054 | 0.143916 | 0.4 |
| 17 | 18 | -0.1285 | -0.08364 | 0.4 |
| 17 | 22 | -0.183 | -0.28678 | 0.4 |
| 18 | 21 | -0.1493 | -0.12693 | 0.2 |
| 18 | 21 | -0.1493 | -0.12693 | 0.2 |
| 19 | 20 | -0.0744 | -0.10516 | 0.2 |
| 19 | 20 | -0.0744 | -0.10516 | 0.2 |
| 20 | 23 | -0.1997 | -0.23044 | 0.3 |
| **21** | **22** | **-0.1999** | **-0.37914** | **0.3** |

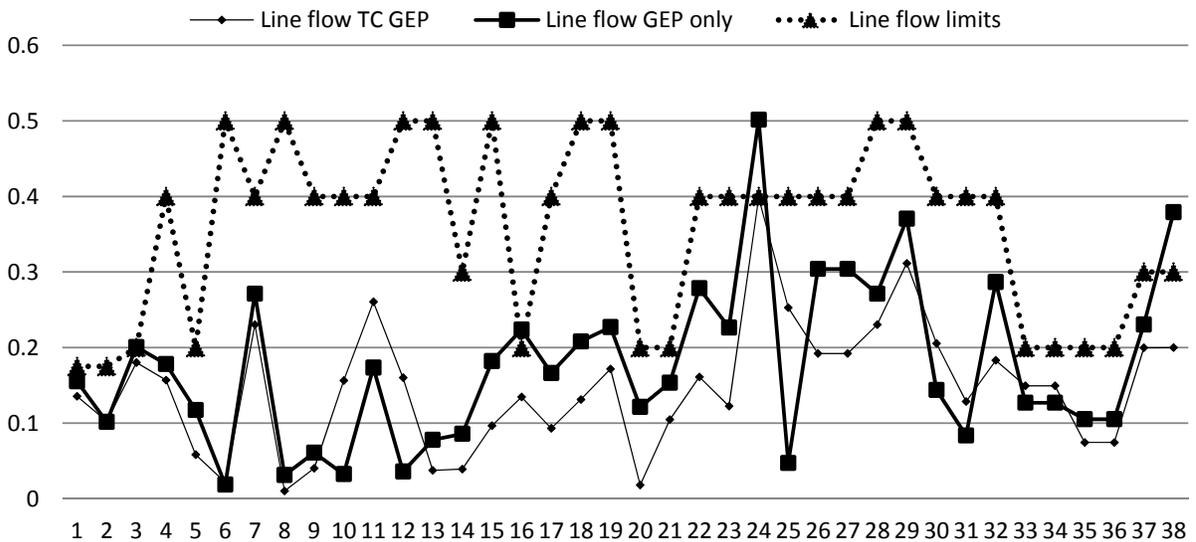

**Fig. 2.5  Line Flow Limits (pu) with TC GEP and simple GEP**

### 2.5.3  DISCUSSION

It can be seen from the result tables shown above that:

1. From the results obtained above we can see that line flows are not within limits in GEP without transmission flow constraints. Hence, TC GEP gives much more practical solution as the power can be evacuated from the generator buses to the load centres with existing transmission capacity.

2. The optimization problem converges after few generations as the fitness value gets saturated.

3. All the constraints are satisfied and the final minimized cost has no penalty value in it.

4. The final results are obtained from multiple executions as GA may prematurely converge sometimes.

5. Monte Carlo Simulation is performed using 400000 samples.

## 2.6    CONCLUSION

In this chapter, Transmission Constrained Generation Expansion Planning has been studied and implemented. The major significance of this problem is the addition of network constraints to the GEP problem.

TC GEP was implemented on IEEE 24 bus RTS using simple Genetic algorithm. DC load flow was used for finding network flows.

GEP without transmission constraints provides a more economical solution than Transmission Constrained GEP as seen from the results in previous section. TC-GEP is 6.64% more costly than simple GEP for 6 years planning horizon.

In spite of this TC GEP is better as it guarantees proper evacuation of power from the generator buses to the load centres without any requirement of transmission network expansion. Here it is considered that sufficient transmission circuits are already available in the system. With GEP without network constraints, the flows can be altered during operations stage by generation rescheduling and other congestion management techniques. This may result in costlier operations which could have been taken care at the planning stage itself. If sufficient Transmission Network corridor is not available we have to perform composite Generation and Transmission Network Expansion Planning.

Monte Carlo Simulations using 400000 samples were performed for validation but the GA gave better solution.

# CHAPTER 3

# COMPOSITE GENERATION AND TRANSMISSION NETWORK EXPANSION PLANNING

## 3.1  INTRODUCTION

In the past decades, much effort has been devoted to the generation expansion planning (GEP) [1–5]. They mainly focused on flexible generation mix under multiobjectives and uncertainties. The locations of generating units and costs of transmission lines are usually neglected. It is customary to assume that there are adequate transmission lines to achieve any generation expansion plans. Actually the practical situations are often different from above premise, especially in some developing countries where transmission networks are very large and weak.

In literature [6], the concept of generic sites was introduced for the purpose of estimating the related transmission costs. Transmission facility costs were added to the capital installation costs of the corresponding generic units to obtain the total cost for each unit.

In this chapter a novel method for composite GEP TNEP is implemented in which both the planning problems are solved simultaneously in a single optimization problem. The objective function is the sum of the investment and operational cost for GEP and the investment cost for the newly added transmission lines. Both static and dynamic models are solved. Dynamic models are used in the evaluation of costs by referring the costs to the first year by adding a discount rate. The string structures for Genetic Algorithm are discussed later in this chapter.

This chapter is organized as follows: The problem of Composite Generation and Transmission Network Expansion Planning is discussed in section (3.2). Section (3.3) deals with the algorithm and solution methodology of the planning problem using Genetic Algorithm. Systems used for implementation and the results of this approach are presented in section (3.5). The last section (3.6) concludes with the important observations.

## 3.2 PROBLEM FORMULATION

The composite GEP and TNEP problem is equivalent to finding a set of best decision vectors over a planning horizon that minimizes the investment and operating costs with several constraints.

The cost objective is represented by the following expression:

$$\min_{U_1 \dots U_T} C = \sum_{t=1}^{T} [I(U_t) + M(X_t) - S(U_t) + I(T_t)] \tag{3.1}$$

where

$$X_t = X_{t-1} + U_t \qquad t = (1,2, \dots T) \tag{3.2}$$

$$I(U_t) = (1+d)^{-2t'} \sum_{i=1}^{N} (CI_i \times U_{t,i}) \tag{3.3}$$

$$S(U_t) = (1+d)^{-2(T+1)} \sum_{i=1}^{N} \left(CI_i \times \delta_i^{2(T-t+1)} \times U_{t,i}\right) \tag{3.4}$$

$$M(X_t) = \sum_{s'=0}^{1} \left( (1+d)^{-(2.5+t'+s')} \times \sum_{i=1}^{N} \left( (X_{t,i} \times FC_i) + MC_i \times EES_{t,i} \right) \right) \tag{3.5}$$

$$I(T_t) = (1+d)^{-2t'} \sum_{l=1}^{\Omega} (c_l \times n_l) \tag{3.6}$$

$$t' = 2(t-1) \ and \ T' = 2 \times T - t'$$

$$U_t = \sum_{i=1}^{N} U_{t,i} \tag{3.7}$$

$$X_t = \sum_{i=1}^{N} X_{t,i} \tag{3.8}$$

$C$       total cost \$;

$U_t$      N-dimensional vector of newly introduced units in the stage t (1 stage =2 years);

$U_{t,i}$      the number of newly introduced units of type $i$ in stage $t$;

$X_t$      cumulative capacity vector of existing units in stage $t$,(MW);

$X_{t,i}$      cumulative capacity of existing units in of type $i$ stage $t$,(MW);

$I(U_t)$      present value of investment cost of the newly introduced unit at the $t^{th}$ stage, \$;

$M(X_t)$      present value of investment cost of the newly introduced unit at the $t^{th}$ stage, \$;

$s'$      variable used to indicate that the maintenance cost is calculated at the middle of each year;

$S(U_t)$      present value of salvage value of the newly added unit at $t^{th}$ interval, \$;

$d$      discount rate;

$CI_i$      capital investment cost of the $i^{th}$ unit, \$;

$\delta_i$      salvage factor of the $i^{th}$ unit;

$T$      length of the planning horizon (in stages);

$N$      total number of different types of units;

$FC$      fixed operation and maintenance cost of the units, \$/MW;

$MC$      variable operation and maintenance cost of the units (energy), \$/MWh;

$EES$      expected energy served, MWh;

$c_l$      cost of line added in lth right-of-way,

$S$      branch-node incidence transposed matrix of the power system,

$f$      vector with elements $f_l$ ,

$\gamma_l$      susceptance of the circuit that can be added to $l^{th}$ right-of-way,

$n_l$      the number of circuits added in $l^{th}$ right-of-way,

$n_l^0$      no. of circuits in the base case,

$\Delta\theta_l$      phase angle difference in $l^{th}$ right-of way,

$f_l$      total real power flow by the circuit in $l^{th}$ right-of-way ,

$\overline{f_l}$      maximum allowed real power flow in the circuit in $l^{th}$ right-of-way,

$\overline{n_i}$  maximum number of circuits that can be added in $l^{th}$ right-of-way,

$\Omega$  set of all right-of-ways,

$nl$  total number of lines in the circuit.

Generation Expansion Constraints are taken as follows:

a) **_Reserve Margin_**_:_ The selected units must satisfy the minimum and maximum reserve margin.

$$(1 + R_{min}) \times D_t \leq \sum_{i=1}^{N} X_{t,i} \leq (1 + R_{max}) \times D_t \tag{3.8}$$

$R_{min}$  minimum reserve margin;

$R_{max}$  maximum reserve margin;

$D_t$  demand at the $t^{th}$ stage in MW

b) **_Upper Construction Limit_**_:_ Let $U_t$ represent the units to be committed in the expansion plan at stage $t$ that must satisfy

$$0 \leq U_t \leq U_{max,t} \tag{3.9}$$

where $U_{max,t}$ is the maximum construction capacity of the units at stage t.

c) **_Demand_**_:_ The selected units must satisfy the demand

$$\sum_{i=1}^{N} X_{t,i} \geq D_t \tag{3.10}$$

d) **_Fuel Mix Ratio_**_:_ The GEP has different types of generating units such as coal, liquefied natural gas (LNG), oil, and nuclear. The selected units along with the existing units of each type must satisfy the fuel mix ratio

$$M_{min}^{j} \leq {X_{t,j}} \Big/ {\sum_{i=1}^{N} X_{t,i}} \leq M_{max}^{j} \tag{3.11}$$

where

$M_{min}^{j}$  minimum fuel mix ratio of $j^{th}$ type;

$M_{max}^{j}$  maximum fuel mix ratio of $j^{th}$ type;

$j$  type of the unit (e.g., oil, LNG, coal, nuclear).

e) **_Reliability criteria_**: The selected units must along with the existing units must satisfy a reliability criterion on loss of load probability

$$LOLP(X_t) \leq \varepsilon \qquad (3.12)$$

where $\varepsilon$ is the reliability criterion for maximum allowable LOLP

Transmission Network Expansion Constraints are taken as follows:

$$S \; f \; + g \; = d \qquad (3.13)$$

$$f_l - \gamma_l \left( n_l^0 + n_l \right)(\Delta \theta_l) = 0, \text{for } l \in 1,2\ldots\ldots, \qquad (3.14)$$

$$\left| f_l \right| \leq \left( n_l^0 + n_l \right) \overline{f_l} \;, \text{for } l \in 1,2\ldots\ldots, nl \qquad (3.15)$$

$$0 \leq n_l \leq \overline{n_l} \qquad (3.16)$$

$f_l$ and $\theta_l$ are unbounded,

$n_l \geq 0,$ and integer, for $l \in 1,2\ldots\ldots, nl,$

$l \in \Omega.$

The objective is to minimize the total investment cost of the new transmission line to be constructed, satisfying the constraint on real power flow in the lines of the network. Constraint (3.13) represents the power balance at each node. Constraint (3.14) is the real power flow equations in DC network. Constraint (3.15) represents the line real power flow constraint. Constraint (3.16) represents the restriction on the construction of lines per corridor (R.O.W). The transmission lines added in any right-of–way are the decision variables.

### 3.3    SOLUTION METHODOLOGY

Simple Genetic Algorithm is used for solving the transmission constrained generation expansion planning problem.

**String Selection:**

| Stage 1 | Stage 2 | ... | ... | Stage t | ... | ... | Stage T-1 | Stage T |
|---------|---------|-----|-----|---------|-----|-----|-----------|---------|

**Sub-string** (Addition of i[th] candidate generator at each stage t and new transmission lines)

**Encoding:** As the range of the variables is between (1-18), 2^5=32,

5 bits are used to represent a candidate transmission network  addition  solution.

Hence, the string will consists of 18X5=40bits if maximum of 18 lines can be added at each stage.

| $U_{t,1}$ | … | … | $U_{t,N-1}$ | $U_{t,N}$ | $T_{t,1}$ | … | … | $T_{t,N-1}$ | $T_{t,N}$ |
|-----------|---|---|-------------|-----------|-----------|---|---|-------------|-----------|

String length for each generator unit i at each generator addition at stage t is taken as 2 bits and for each new transmission lines string length is taken as 5 bits.

Therefore, at each stage for 21 different generator units each substring length becomes 2×21 = 42 and for the 18 new transmission lines that can be added in each stage, bits required is 5×18=90. Hence total sub-string length is 132.

For 3 stages or 6 year planning horizon string length becomes 3×132 = 396

**Fitness Function:**

$$fitness = \frac{\alpha}{1+J} \tag{3.17}$$

Where α is a constant, J is the objective function and penalties for any constraint violation

**Parameters for GEP:**

$d = $ *discount rate* = 8.5%

*LOLP criteria* = 0.01

*Min. Reserve Margin* = 20% *Max. Reserve Margin* = 60%

*Parent Selection Method:* **Tournament Selection**

*Crossover Technique:* **Uniform Crossover (because of the large string length)**

### Table 3.1 Genetic Algorithm Parameters

| GA Parameters | Value |
|---------------|-------|
| **Population Size** | 200 |
| **Maximum Generations / Iterations** | 2000 |
| **Probabilities of Crossover and Mutation** | 0.9, 0.01 |
| **Number of Elite Strings** | 5 |

The modified IEEE 24 bus RTS used is formed by deliberately removing some crucial lines from the original topology so that there is transmission network inefficiency. The algorithm will add the lines as required for the final topology to become feasible.

## 3.4     RESULTS

This section presents the results obtained after the implementation of the Composite GEP and TNEP problem using Genetic Algorithm. The modified IEEE 24 bus Reliability Test System (RTS) has been used for implementation of the problem.

### 3.4.1   IEEE 24 BUS RTS SYSTEM

**Table 3.2 Technical and Economic Data of the Existing Plants**

| Name (Fuel Type) | No. of Units | Unit Capacity (MW) | FOR (%) | Operating Cost ($/KWh) | Fixed Cost (S/KW-Month) |
|---|---|---|---|---|---|
| LNG #1 | 1 | 591 | 7 | 0.024 | 2.25 |
| Oil #1 | 1 | 40 | 3 | 0.043 | 4.52 |
| Oil #2 | 1 | 40 | 10 | 0.038 | 1.63 |
| Oil #3 | 1 | 300 | 10 | 0.04 | 1.63 |
| Coal #1 | 1 | 152 | 15 | 0.023 | 6.65 |
| Coal #2 | 1 | 152 | 9 | 0.019 | 2.81 |
| Coal #3 | 1 | 155 | 8.5 | 0.015 | 2.81 |
| Coal #4 | 1 | 155 | 8.5 | 0.015 | 2.81 |
| Coal #5 | 1 | 300 | 8.5 | 0.015 | 2.81 |
| Coal #6 | 1 | 660 | 8.5 | 0.015 | 2.81 |
| Nuclear #1 | 1 | 400 | 9 | 0.005 | 4.94 |
| Nuclear #2 | 1 | 400 | 8.8 | 0.005 | 4.63 |

### Table 3.3 Technical and Economic Data of the Candidate Plants

| Candidate Type | Construction Upper Limit | Unit Capacity (MW) | FOR(%) | Operating Cost ($/KWh) | Fixed Cost (S/KW-Month) | Capital Cost ($/KW) | Life Time (Yrs) |
|---|---|---|---|---|---|---|---|
| LNG #1 | 5 | 100 | 7 | 0.021 | 2.2 | 812.5 | 25 |
| LNG #2 | 3 | 100 | 7 | 0.021 | 2.2 | 812.5 | 25 |
| LNG #3 | 2 | 50 | 7 | 0.021 | 2.2 | 812.5 | 25 |
| LNG #4 | 5 | 200 | 7 | 0.021 | 2.2 | 812.5 | 25 |
| LNG #5 | 6 | 200 | 7 | 0.021 | 2.2 | 812.5 | 25 |
| LNG #6 | 6 | 200 | 7 | 0.021 | 2.2 | 812.5 | 25 |
| Oil #1 | 3 | 100 | 10 | 0.035 | 0.9 | 500 | 20 |
| Oil #2 | 5 | 100 | 10 | 0.035 | 0.9 | 500 | 20 |
| Oil #3 | 4 | 50 | 10 | 0.035 | 0.9 | 500 | 20 |
| Oil #4 | 3 | 100 | 10 | 0.035 | 0.9 | 500 | 20 |
| Coal #1 | 5 | 100 | 9.5 | 0.014 | 2.75 | 1062.5 | 25 |
| Coal #2 | 5 | 100 | 9.5 | 0.014 | 2.75 | 1062.5 | 25 |
| Coal #3 | 5 | 100 | 9.5 | 0.014 | 2.75 | 1062.5 | 25 |
| Coal #4 | 6 | 200 | 9.5 | 0.014 | 2.75 | 1062.5 | 25 |
| Coal #5 | 6 | 200 | 9.5 | 0.014 | 2.75 | 1062.5 | 25 |
| Coal #6 | 5 | 100 | 9.5 | 0.014 | 2.75 | 1062.5 | 25 |
| Coal #7 | 4 | 200 | 9.5 | 0.014 | 2.75 | 1062.5 | 25 |
| Nuclear #1 | 3 | 100 | 9 | 0.004 | 4.6 | 1625 | 25 |
| Nuclear #2 | 5 | 100 | 9 | 0.004 | 4.6 | 1625 | 25 |
| Nuclear #3 | 5 | 100 | 9 | 0.004 | 4.6 | 1625 | 25 |
| Nuclear #4 | 5 | 100 | 9 | 0.004 | 4.6 | 1625 | 25 |

**Table 3.4 Technical and Economic Data of the Candidate Plants**

| Line # | From Bus | To Bus | Capacity (MW) | Cost(k$/MW) |
|--------|----------|--------|---------------|-------------|
| 1 | 1 | 2 | 175 | 43.9 |
| 2 | 1 | 3 | 175 | 80.5 |
| 3 | 3 | 24 | 500 | 21.5 |
| 4 | 7 | 9 | 500 | 73.1 |
| 5 | 8 | 9 | 500 | 62.9 |
| 6 | 9 | 11 | 500 | 21.5 |
| 7 | 10 | 12 | 500 | 21.5 |
| 8 | 11 | 13 | 500 | 21.85 |
| 9 | 11 | 14 | 500 | 19.2 |
| 10 | 12 | 23 | 800 | 42 |
| 11 | 14 | 16 | 500 | 18 |
| 12 | 15 | 21 | 1000 | 22.5 |
| 13 | 15 | 24 | 500 | 23.4 |
| 14 | 16 | 17 | 500 | 12.6 |
| 15 | 17 | 22 | 500 | 48.3 |
| 16 | 18 | 21 | 500 | 11.9 |
| 17 | 19 | 20 | 500 | 18.2 |
| 18 | 20 | 23 | 500 | 9.93 |

### 3.4.2 Composite GEP TNEP Problem - Single Stage Static Problem (2 year Planning Horizon)

| Composite GEP TNEP Problem - Single Stage Static Problem |
|---|
| Minimized Total Cost:  $1.3343 \times 10^9$ $ |
| Cost for TNEP: $2.56386 \times 10^7$ $ |
| Cost for GEP: $1.3087 \times 10^9$ $ |

**Table 3.5-3.6:** Capacity addition of new units of candidate generators and transmission lines

| Unit# | Stage 1 Units Added |
|---|---|
| LNG #1 | 0 |
| LNG #2 | 1 |
| LNG #3 | 0 |
| LNG #4 | 0 |
| LNG #5 | 2 |
| LNG #6 | 2 |
| Oil #1 | 2 |
| Oil #2 | 0 |
| Oil #3 | 1 |
| Oil #4 | 1 |
| Coal #1 | 0 |
| Coal #2 | 0 |
| Coal #3 | 0 |
| Coal #4 | 0 |
| Coal #5 | 0 |
| Coal #6 | 0 |
| Coal #7 | 1 |
| Nuclear #1 | 0 |
| Nuclear #2 | 0 |
| Nuclear #3 | 0 |
| Nuclear #4 | 0 |

| Line # | From Bus | To Bus | Lines Added (Stage 1) |
|---|---|---|---|
| 1 | 1 | 2 | 1 |
| 2 | 1 | 3 | 0 |
| 3 | 3 | 24 | 0 |
| 4 | 7 | 9 | 0 |
| 5 | 8 | 9 | 0 |
| 6 | 9 | 11 | 0 |
| 7 | 10 | 12 | 0 |
| 8 | 11 | 13 | 0 |
| 9 | 11 | 14 | 0 |
| 10 | 12 | 23 | 0 |
| 11 | 14 | 16 | 0 |
| 12 | 15 | 21 | 1 |
| 13 | 15 | 24 | 0 |
| 14 | 16 | 17 | 0 |
| 15 | 17 | 22 | 0 |
| 16 | 18 | 21 | 0 |
| 17 | 19 | 20 | 0 |
| 18 | 20 | 23 | 0 |

### 3.4.3 Separate GEP and TNEP Problem - Single Stage Static Problem (2 year Planning Horizon)

Initially, simple GEP is performed to find the new generator units and then TNEP is performed so that power can be evacuated to the load centers.

| Separate GEP and TNEP Problem - Single Stage Static Problem |
|---|
| Minimized Total Cost:   $1.3393 \times 10^9$ $ |
| Minimized Cost for GEP: $1.3087 \times 10^9$ $ |
| Minimized Cost for TNEP: $3.0693 \times 10^7$ $ |

**Table 3.7-3.8:** Capacity addition of new units of candidate generators and transmission lines

| Unit# | Stage 1 Units Added |
|---|---|
| LNG #1 | 0 |
| LNG #2 | 1 |
| LNG #3 | 0 |
| LNG #4 | 2 |
| LNG #5 | 2 |
| LNG #6 | 0 |
| Oil #1 | 2 |
| Oil #2 | 0 |
| Oil #3 | 1 |
| Oil #4 | 1 |
| Coal #1 | 0 |
| Coal #2 | 0 |
| Coal #3 | 0 |
| Coal #4 | 0 |
| Coal #5 | 0 |
| Coal #6 | 0 |
| Coal #7 | 1 |
| Nuclear #1 | 0 |
| Nuclear #2 | 0 |
| Nuclear #3 | 0 |
| Nuclear #4 | 0 |

| Line # | From Bus | To Bus | Lines Added (Stage 1) |
|---|---|---|---|
| 1 | 1 | 2 | 1 |
| 2 | 1 | 3 | 0 |
| 3 | 3 | 24 | 0 |
| 4 | 7 | 9 | 0 |
| 5 | 8 | 9 | 0 |
| 6 | 9 | 11 | 0 |
| 7 | 10 | 12 | 0 |
| 8 | 11 | 13 | 0 |
| 9 | 11 | 14 | 0 |
| 10 | 12 | 23 | 0 |
| 11 | 14 | 16 | 0 |
| 12 | 15 | 21 | 1 |
| 13 | 15 | 24 | 0 |
| 14 | 16 | 17 | 0 |
| 15 | 17 | 22 | 0 |
| 16 | 18 | 21 | 1 |
| 17 | 19 | 20 | 0 |
| 18 | 20 | 23 | 0 |

### 3.4.4 Composite GEP TNEP Dynamic Problem (6 Year Planning Horizon)

| Composite GEP TNEP Dynamic Problem |
|:---:|
| **Minimized Total Cost:**    3.6368×10⁹ $ |
| **Cost of new transmission lines:**  0.1143×10⁹ $ |
| **Cost of new Generator Units:**  3.5225×10⁹ $ |

**Table 3.9 Stage-wise capacity addition of new units of candidate plants**

| SL. No. | Stage 1 | Stage 2 | Stage 3 | Total |
|:---:|:---:|:---:|:---:|:---:|
| **LNG #1** | 0 | 0 | 2 | 2 |
| **LNG #2** | 1 | 1 | 0 | 2 |
| **LNG #3** | 0 | 0 | 0 | 0 |
| **LNG #4** | 0 | 0 | 0 | 0 |
| **LNG #5** | 0 | 0 | 0 | 0 |
| **LNG #6** | 2 | 0 | 0 | 2 |
| **Oil #1** | 0 | 1 | 2 | 3 |
| **Oil #2** | 2 | 0 | 0 | 2 |
| **Oil #3** | 1 | 0 | 0 | 1 |
| **Oil #4** | 2 | 1 | 0 | 3 |
| **Coal #1** | 0 | 0 | 2 | 2 |
| **Coal #2** | 2 | 2 | 0 | 4 |
| **Coal #3** | 0 | 0 | 0 | 0 |
| **Coal #4** | 0 | 0 | 2 | 2 |
| **Coal #5** | 0 | 0 | 0 | 0 |
| **Coal #6** | 0 | 0 | 0 | 0 |
| **Coal #7** | 1 | 0 | 0 | 1 |
| **Nuclear #1** | 1 | 1 | 0 | 2 |
| **Nuclear #2** | 0 | 0 | 0 | 0 |
| **Nuclear #3** | 0 | 0 | 0 | 0 |
| **Nuclear #4** | 0 | 0 | 1 | 1 |

**Table 3.10 Stage-wise capacity addition of new Transmission Lines**

| Line # | From Bus | To Bus | Lines Added (Stage 1) | Lines Added (Stage 2) | Lines Added (Stage 3) |
|---|---|---|---|---|---|
| 1 | 1 | 2 | 0 | 0 | 0 |
| 2 | 1 | 3 | **1** | 0 | 0 |
| 3 | 3 | 24 | **1** | 0 | 0 |
| 4 | 7 | 9 | 0 | 0 | 0 |
| 5 | 8 | 9 | 0 | 0 | 0 |
| 6 | 9 | 11 | **2** | 0 | 0 |
| 7 | 10 | 12 | **1** | **1** | 0 |
| 8 | 11 | 13 | 0 | 0 | 0 |
| 9 | 11 | 14 | 0 | 1 | **1** |
| 10 | 12 | 23 | 0 | 0 | 0 |
| 11 | 14 | 16 | 0 | **1** | **2** |
| 12 | 15 | 21 | 0 | 0 | 0 |
| 13 | 15 | 24 | 0 | **1** | 0 |
| 14 | 16 | 17 | 0 | 0 | **2** |
| 15 | 17 | 22 | 0 | 0 | 0 |
| 16 | 18 | 21 | 0 | **1** | 0 |
| 17 | 19 | 20 | 0 | 0 | 0 |
| 18 | 20 | 23 | **1** | 0 | **1** |

**Table 3.11 Generation and Reserves at Each Stage**

| Stage | Peak MW Demand at each stage | Total Generation Capacity at each stage (MW) | Reserve at each stage (MW) |
|---|---|---|---|
| 0 | 3238 | 3345 | 107 |
| 1 | 3885.63 | 4595 | 709.4 |
| 2 | 4662.72 | 5395 | 732.3 |
| 3 | 5595.3 | 6645 | 1049.7 |

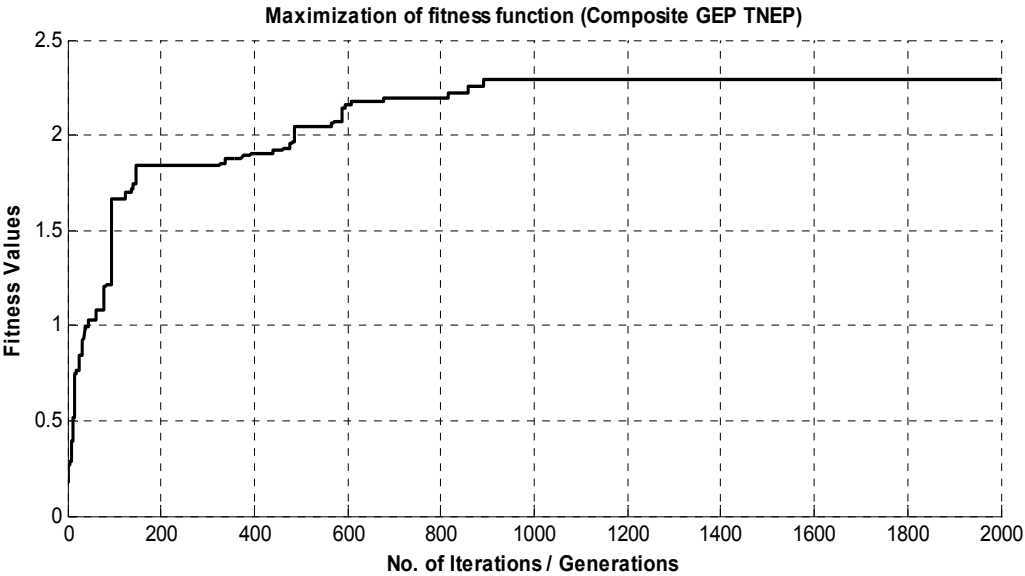

**Fig 3.1 Maximization of Fitness Function**

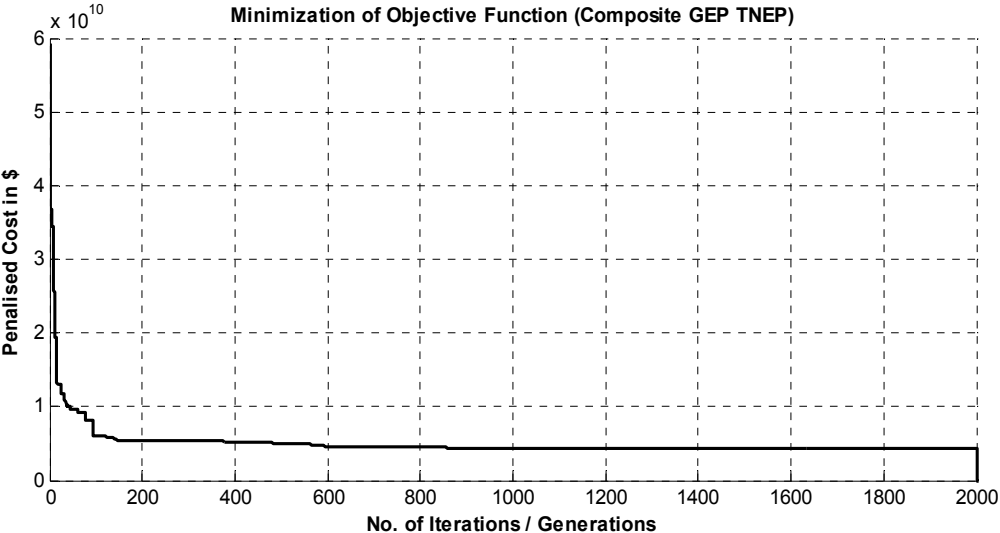

**Fig 3.2  Minimization of Objective function**

### 3.4.5 Discussions

It can be seen from the result tables shown above that:

1. We can observe that the composite GEP TNEP gives 16.47% economical solution than separate generation and transmission expansion planning.

2. The optimization problem converges after few generations as the fitness value saturates.

3. All the constraints are satisfied and the final minimized cost has no penalty value in it.

4. The final results are obtained from multiple executions as GA may prematurely converge sometimes.

5. Monte Carlo Simulation is performed using 400000 samples for a single stage problem for validation.

### 3.5     CONCLUSIONS

In this chapter, a novel approach for composite Generation and Transmission Network Expansion Planning is implemented by a single optimization process. The main significance of this method is the novel technique by which GEP and TNEP were integrated.

The algorithm was implemented on IEEE 24 bus RTS using simple Genetic Algorithm. DC load flow was used for finding network flows.

From the results it can be shown that for a single stage problem, composite GEP TNEP provided more economical solution with 16.47% decrease in the cost for TNEP than in separate GEP and TNEP approach.

In the previous chapter it was observed that TC-GEP was better than normal GEP as the former guaranteed power evacuation from generator buses to load centres. But if sufficient transmission circuit is unavailable, TNEP has to be performed. TNEP can be done after GEP but it may not be economical. As is observed from the results above for composite GEP TNEP the generation expansion cost is less compared to separate GEP TNEP as some of the loads can be fed just by enhancing the transmission network or by just modifying the network topology without any capacity addition. Thus composite GEP TNEP provides better planning than the traditional separate expansion method.

Monte Carlo Simulations using 400000 samples were performed for validation but the GA gave better solution.

# CHAPTER 4

# TRANSMISSION NETWORK EXPANSION PLANNING (TNEP) USING AC MODEL

## 4.1    INTRODUCTION

The objective of the power system transmission network expansion planning (TNEP) problem is to determine 'where', 'how many' and 'when' new devices must be added to a network in order to make its operation viable for a pre-defined horizon of planning, at a minimum cost (determine the optimum expansion plan). The system network of the base year, the candidate circuits, the power generation and power demand of the planning horizon and investment constraints are the basic data for the problem.

In literature [36], TNEP using AC model is performed using a constructive heuristic method. Traditionally, TNEP is modelled by simple DC model to add lines required for real power transfer and later system is reinforced by reactive power planning. Some of the disadvantages of TNEP by DC model and advantages of using AC model is discussed below.

The use of the DC model for TNEP has the following disadvantages, among others:

(1) the transmission expansion planning problem must be separated from the reactive power planning problem;

(2) it is frequently necessary to reinforce an expansion plan obtained using the DC model, when an operation with the AC model is considered; and

(3) the difficulty of taking into account the power losses in the initial phase of the planning

The use of the AC model in TNEP has following advantages:

(1) efficiently carrying out the reinforcement stage, when the expansion plans obtained for the DC model are used in the reactive planning phase;

(2) using an integrated mathematical model that allows transmission network expansion planning problems and the optimal allocation of reactive power (simultaneously, in a unique stage), dispensing the use of simplified models such as the DC model;

(3) incorporating the determination of the transmission system's precise real losses in a trivial way and as a sub-product of the optimisation process;

(4) incorporating other nonlinear operation characteristic devices in the TNEP problem, for example, the FACTS controllers; and

(5) the possibility of carrying out other types of studies, after solving the AC integrated TNEP problem, for example: voltage stability, nodal analysis, transient stability analysis and so on.

Moreover, the difficulties that appear when working with planning using the AC model are:

(a) handling disconnected systems, a common situation in the initial phase of transmission planning, when generators and loads have not yet been electrically connected to the network;

(b) developing an efficient optimisation technique; and

(c) working with large nonlinear programming problems.

In this chapter, AC model based TNEP is solved by Genetic Algorithm and Fast Decoupled Load Flow technique was used.

This chapter is organized as follows: The problem formulation Transmission Network Expansion Planning using AC model is discussed in section (4.2). Section (4.3) deals with the algorithm and solution methodology of the planning problem using Genetic Algorithm. Systems used for implementation and the results of this approach are presented in section (4.5). The last section (4.6) concludes with the important observations.

## 4.2    PROBLEM FORMULATION

$$Min\ v = c^T n \qquad\qquad (4.1)$$

Subject to

$$P(v, \theta, n) - P_G + P_D = 0 \qquad\qquad (4.2)$$

$$Q(v, \theta, n) - Q_G + Q_D = 0 \qquad\qquad (4.3)$$

$$\underline{P_G} \leq P_G \leq \overline{P_G} \qquad\qquad (4.4)$$

$$\underline{Q_G} \leq Q_G \leq \overline{Q_G} \qquad\qquad (4.5)$$

$$\underline{V} \leq V \leq \overline{V} \qquad\qquad (4.6)$$

$$(N + N_0)S^{from} \leq (N + N_0)\overline{S} \qquad\qquad (4.7)$$

$$(N + N_0)S^{to} \leq (N + N_0)\overline{S} \qquad\qquad (4.8)$$

$$0 \leq n \leq \underline{n} \qquad\qquad (4.9)$$

$n$ integer and $\theta$ unbounded

Security Constraints is an important part of TNEP. Here, **N-1** contingency analysis is used to check the system security. The **N-1** contingency analysis looks at the system state after a single line outage. Results for both with and without Security Constraints are carried out.

Where,

$c$ **:** circuit cost vector that can be added to the network

$n$ **:** added circuit vector, respectively.

$N$, $N_0$ : diagonal matrices containing vector $n$ and the existing circuits in the base configuration, respectively.

$v$ : investment due to the addition of circuits to the networks.

$\overline{n}$ : vector containing the maximum number of circuits that can be added.

$\theta$ : phase angle vector.

$P_G$ , $Q_G$ : real and reactive power generation vectors.

$P_D$, $Q_D$ : real and reactive power demand vectors;

$V$ is the voltage magnitude vector;

$\overline{P_G}$, $\overline{Q_G}$, $\overline{V}$ : vectors of maximum limits of generation of real power, reactive power and voltage magnitudes,

$\underline{P_G}$, $\underline{Q_G}$, $\underline{V}$, : vectors of minimum limits of generation of real power, reactive power and voltage magnitudes,

$S^{from}$, $S^{to}$, $\overline{S}$ : apparent power flow vectors (MVA) in the branches in both terminals and their limits.

The elements of vector $\boldsymbol{P(v, \theta, n)}$ and $\boldsymbol{Q(v, \theta, n)}$ are calculated by

$$P_i(v, \theta, n) = V_i \sum_{j \in N} V_j [G_{ij}(n) \cos \theta_{ij} + B_{ij}(n) \sin \theta_{ij}] \qquad (4.10)$$

$$Q_i(v, \theta, n) = V_i \sum_{j \in N} V_j [G_{ij}(n) \sin \theta_{ij} - B_{ij}(n) \cos \theta_{ij}] \qquad (4.11)$$

where i, $\boldsymbol{j \in N}$ represent the buses and N is the set of all buses, *ij* represents the circuit between buses *i* and *j*. The bus admittance matrix elements are

$$G = \begin{cases} G_{ij}(n) = -(n_{ij}g_{ij} + n^0{}_{ij}g^0{}_{ij}) \\ G_{ii}(n) = \sum_{j \in \Omega_i} (n_{ij}g_{ij} + n^0{}_{ij}g^0{}_{ij}) \end{cases}$$

$$(4.12)$$

$$B = \begin{cases} B_{ij}(n) = -(n_{ij}b_{ij} + n^0{}_{ij}b^0{}_{ij}) \\ B_{ii}(n) = b^{sh}{}_i + \sum_{j \in \Omega_i} \begin{aligned} &[n_{ij}(b_{ij} + b^{sh}{}_{ij}) \\ &+ n^0{}_{ij}(b^0{}_{ij} + (b^{sh}{}_{ij})^0)] \end{aligned} \end{cases}$$

where $\boldsymbol{\Omega_i}$ represents the set of all buses directly connected to bus $\boldsymbol{i}$; $g_{ij}$ , $b_{ij}$ and $b_{ij}{}^{sh}$ are the conductance, susceptance and shunt susceptance of the transmission line or transformerij (if ij is a transformer $b_{ij}{}^{sh} = 0$), respectively, and $b_i{}^{sh}$ is the shunt susceptance at bus $i.$ Elements (ij) of vectors S ($S^{from}$, $S^{to}$) are given by

$$S_{ij} = \sqrt{P_{ij}{}^2 + Q_{ij}{}^2} \tag{4.13}$$

$$P_{ij} = V_j^2 g_{ij} - V_i V_j (g_{ij} \cos \theta_{ij} - b_{ij} \sin \theta_{ij}) \tag{4.14}$$

$$Q_{ij} = -V_j^2 (b_{ij} + b^{sh}{}_{ij}) + V_i V_j (g_{ij} \sin \theta_{ij} + b_{ij} \cos \theta_{ij}) \tag{4.15}$$

## 4.3     SOLUTION METHODOLOGY

Simple Genetic Algorithm is used for solving the TNEP using AC model.

For Garver's 6bus system :

Bits required for each Transmission line to be added is 4 as there are 15 possible lines for Garver's 6 bus system ($2^4 = 16$)

Therefore, for 15 lines that can be added at max, string length becomes 15×4=60

**Fitness Function:**

$$fitness = \frac{\alpha}{1+J} \tag{2.21}$$

Where α is a constant, J is the sum of the objective function and penalties for any constraint violation

**Parent Selection Method:** Tournament Selection

**Crossover Technique:** Uniform Crossover

### Table 4.1 Genetic Algorithm Parameters

| GA Parameters | Value |
|---|---|
| **Population Size** | 80 |
| **Maximum Generations / Iterations** | 200 |
| **Probabilities of Crossover and Mutation** | 0.9, 0.01 |
| **Number of Elite Strings** | 5 |

For the implementation of TNEP using AC model load is assumed to have a constant power factor of 0.9 and the optimization is done for three load scenarios, viz. 22.5% above base case for duration of 876 hrs, base case for 6132 hrs and 30 % below base case for 1752 hrs for an yearly load duration curve. Fast decoupled Load Flow (FDLF) is used for evaluation of the state variables and hence the injections, flows and losses.

**Table 4.2 Load Scenarios**

| Load | Duration (Hrs) |
|---|---|
| **22.5% above base case** | **876** |
| **Base Case** | **6132** |
| **30% below base case** | **1752** |

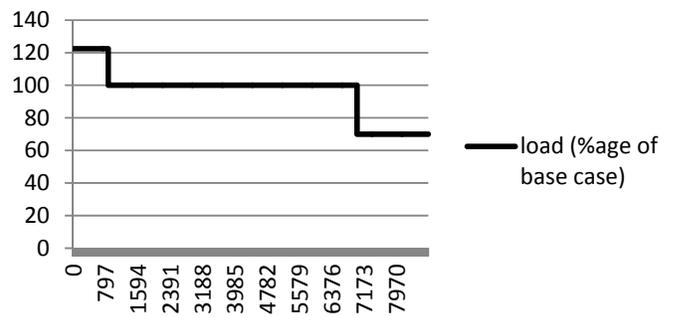

**4.4    RESULTS**

This section presents the results obtained after the implementation of the TNEP using AC model problem using Genetic Algorithm. The Garver's 6 bus system has been used for implementation of the problem.

**Table 4.3 Technical and Economic Data of the candidate lines**

| Line # | From Bus | To Bus | Capacity (MVA) | Cost(million $) |
|--------|----------|--------|----------------|-----------------|
| 1 | 1 | 2 | 120 | 40 |
| 2 | 1 | 3 | 120 | 38 |
| 3 | 1 | 4 | 100 | 60 |
| 4 | 1 | 5 | 120 | 20 |
| 5 | 1 | 6 | 90 | 68 |
| 6 | 2 | 3 | 120 | 20 |
| 7 | 2 | 4 | 120 | 40 |
| 8 | 2 | 5 | 120 | 31 |
| 9 | 6 | 2 | 120 | 30 |
| 10 | 3 | 4 | 120 | 59 |
| 11 | 3 | 5 | 120 | 20 |
| 12 | 6 | 3 | 120 | 48 |
| 13 | 4 | 5 | 95 | 63 |
| 14 | 4 | 6 | 120 | 30 |
| 15 | 5 | 6 | 98 | 61 |

---

**AC TNEP without Security Constraints for Garver's 6 Bus Test System**

**Minimized Total Cost without Security Constraints: 3.11×10[8] $**

---

**Table 4.4 Addition of lines for the given load scenarios and pf 0.9 for Garver's 6 bus system by AC TNEP without Security Constraints**

| Line # | From Bus | To Bus | Lines Present | Lines Added |
|--------|----------|--------|---------------|-------------|
| 1 | 1 | 2 | 1 | 0 |
| 2 | 1 | 3 | 0 | 0 |
| 3 | 1 | 4 | 1 | 0 |
| 4 | 1 | 5 | 1 | 0 |
| 5 | 1 | 6 | 0 | 0 |
| 6 | 2 | 3 | 1 | 0 |
| 7 | 2 | 4 | 1 | 0 |
| 8 | 2 | 5 | 0 | 0 |
| 9 | 6 | 2 | 1 | 4 |
| 10 | 3 | 4 | 0 | 0 |
| 11 | 3 | 5 | 1 | 2 |
| 12 | 6 | 3 | 0 | 0 |
| 13 | 4 | 5 | 0 | 0 |
| 14 | 4 | 6 | 0 | 3 |
| 15 | 5 | 6 | 0 | 1 |

**Table 4.5 AC Load flow results after AC TNEP without Security Constraints**

| Bus | Voltage pu | Angle deg | Pgen pu | Qgen pu |
|-----|-----------|-----------|---------|---------|
| 1 | 1.04 | 0 | 0.108 | 0.1 |
| 2 | 1.0342 | 0.047 | 0 | 0 |
| 3 | 1.04 | 0.332 | 0.247 | 0.133 |
| 4 | 1.0325 | 0.044 | 0 | 0 |
| 5 | 1.0337 | -0.267 | 0 | 0 |
| 6 | 1.04 | 0.862 | 0.407 | 0.146 |

**Table 4.6 Line flows after AC TNEP without Security Constraints**

| From Bus | To Bus | Real Power flow (pu) | Reactive Power Flow (pu) | Apparent Power Flow (pu) | Flow Limit (pu) |
|---|---|---|---|---|---|
| 1 | 2 | 0.011 | 0.0141 | 0.0179 | 0.12 |
| 1 | 4 | 0.0108 | 0.0119 | 0.0161 | 0.10 |
| 1 | 5 | 0.0485 | 0.0283 | 0.0562 | 0.12 |
| 2 | 3 | -0.0277 | -0.0274 | 0.0389 | 0.12 |
| 2 | 4 | 0.0051 | 0.0038 | 0.0063 | 0.12 |
| 6 | 2 | 0.0413 | 0.0163 | 0.0444 | 0.12 |
| 6 | 2 | 0.0413 | 0.0163 | 0.0444 | 0.12 |
| 6 | 2 | 0.0413 | 0.0163 | 0.0444 | 0.12 |
| 6 | 2 | 0.0413 | 0.0163 | 0.0444 | 0.12 |
| 6 | 2 | 0.0413 | 0.0163 | 0.0444 | 0.12 |
| 3 | 5 | 0.0542 | 0.0277 | 0.0609 | 0.12 |
| 3 | 5 | 0.0542 | 0.0277 | 0.0609 | 0.12 |
| 3 | 5 | 0.0542 | 0.0277 | 0.0609 | 0.12 |
| 4 | 6 | -0.048 | -0.0207 | 0.0523 | 0.12 |
| 4 | 6 | -0.048 | -0.0207 | 0.0523 | 0.12 |
| 4 | 6 | -0.048 | -0.0207 | 0.0523 | 0.12 |
| 5 | 6 | -0.029 | -0.0076 | 0.0299 | 0.098 |

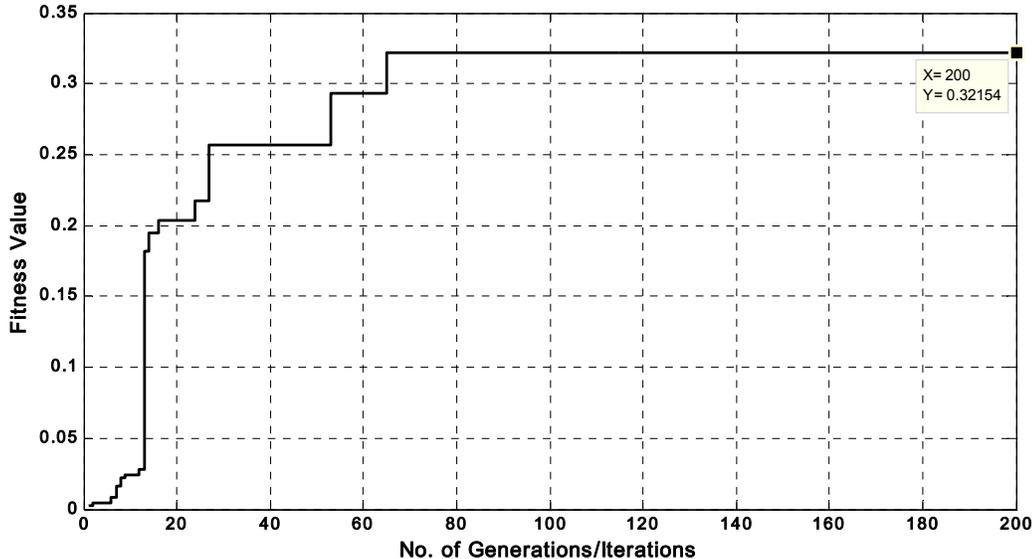

**Fig 4.1 Maximization of Fitness Function for TNEP AC model without Security Constraints**

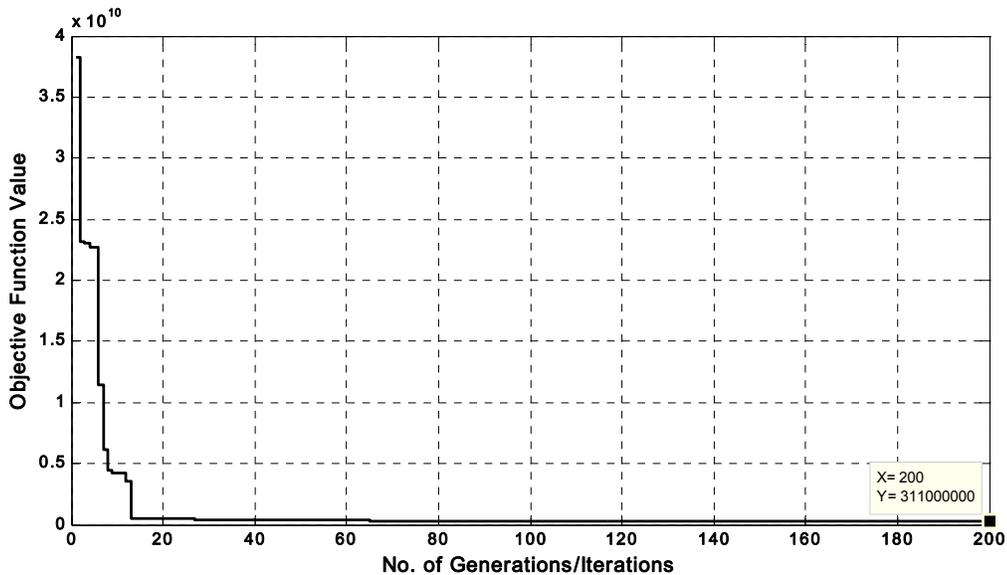

**Fig 4.2 Minimization of total investment cost for TNEP AC model without Security Constraints**

| AC TNEP with Security Constraints for Garver's 6 Bus Test System |
|:---:|
| Minimized Total Cost with Security Constraints: $3.49 \times 10^8$ $ |

Table 4.7   Addition of lines for the given load scenarios and pf 0.9 for
Garver's 6 bus system by AC TNEP with Security Constraints

| Line # | From Bus | To Bus | Lines Present | Lines Added |
|:---:|:---:|:---:|:---:|:---:|
| 1 | 1 | 2 | 1 | 0 |
| 2 | 1 | 3 | 0 | **1** |
| 3 | 1 | 4 | 1 | 0 |
| 4 | 1 | 5 | 1 | 0 |
| 5 | 1 | 6 | 0 | 0 |
| 6 | 2 | 3 | 1 | 0 |
| 7 | 2 | 4 | 1 | 0 |
| 8 | 2 | 5 | 0 | 0 |
| 9 | 6 | 2 | 1 | **4** |
| 10 | 3 | 4 | 0 | 0 |
| 11 | 3 | 5 | 1 | **2** |
| 12 | 6 | 3 | 0 | 0 |
| 13 | 4 | 5 | 0 | 0 |
| 14 | 4 | 6 | 0 | **3** |
| 15 | 5 | 6 | 0 | **1** |

**Table 4.8 AC Load flow results after AC TNEP with Security Constraints**

| Bus | Voltage pu | Angle deg | Pgen pu | Qgen pu |
|-----|------------|-----------|---------|---------|
| 1 | 1.04 | 0 | 0.107 | 0.097 |
| 2 | 1.034 | 0.023 | 0 | 0 |
| 3 | 1.04 | 0.238 | 0.247 | 0.129 |
| 4 | 1.033 | -0.071 | 0 | 0 |
| 5 | 1.034 | -0.344 | 0 | 0 |
| 6 | 1.04 | 0.705 | 0.407 | 0.152 |

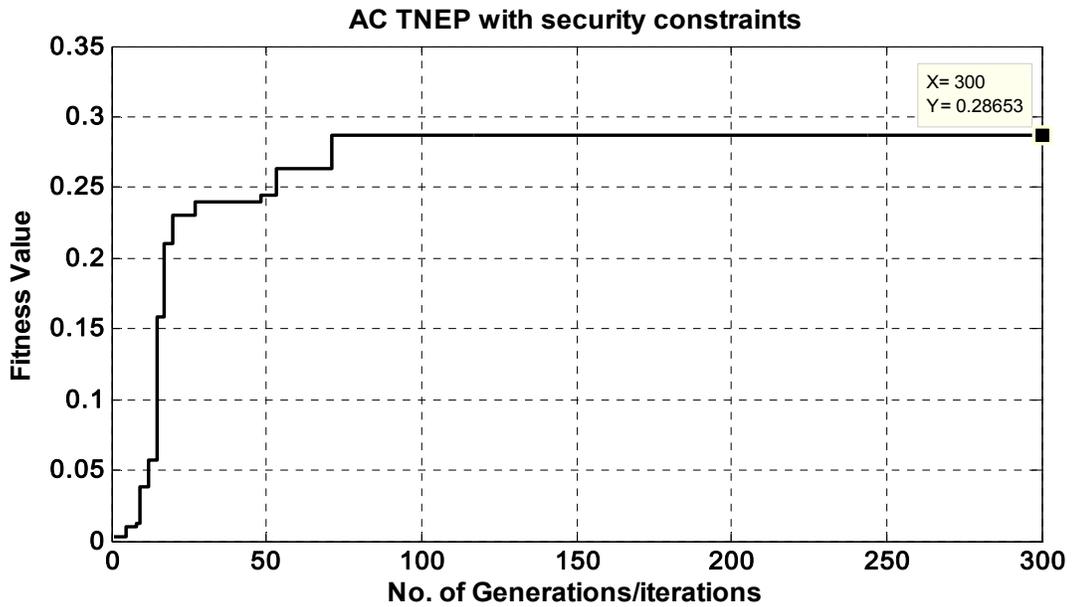

**Fig 4.3 Maximization of Fitness Function for TNEP AC model with Security Constraints**

**Table 4.9 Line flows after AC TNEP with Security Constraints**

| From Bus | To Bus | Real Power flow (pu) | Reactive Power Flow (pu) | Apparent Power Flow (pu) | Flow Limit (pu) |
|---|---|---|---|---|---|
| 1 | 2 | 0 | 0.015 | 0.015 | 0.12 |
| 1 | 3 | -0.012 | 0.001 | 0.012 | 0.12 |
| 1 | 4 | 0.003 | 0.013 | 0.013 | 0.10 |
| 1 | 5 | 0.035 | 0.029 | 0.045 | 0.12 |
| 2 | 3 | -0.023 | -0.028 | 0.036 | 0.12 |
| 2 | 4 | 0.005 | 0.004 | 0.006 | 0.12 |
| 6 | 2 | 0.044 | 0.016 | 0.047 | 0.12 |
| 6 | 2 | 0.044 | 0.016 | 0.047 | 0.12 |
| 6 | 2 | 0.044 | 0.016 | 0.047 | 0.12 |
| 6 | 2 | 0.044 | 0.016 | 0.047 | 0.12 |
| 6 | 2 | 0.044 | 0.016 | 0.047 | 0.12 |
| 3 | 5 | 0.057 | 0.027 | 0.063 | 0.12 |
| 3 | 5 | 0.057 | 0.027 | 0.063 | 0.12 |
| 3 | 5 | 0.057 | 0.027 | 0.063 | 0.12 |
| 4 | 6 | -0.051 | -0.02 | 0.055 | 0.12 |
| 4 | 6 | -0.051 | -0.02 | 0.055 | 0.12 |
| 4 | 6 | -0.051 | -0.02 | 0.055 | 0.12 |
| 5 | 6 | -0.033 | -0.007 | 0.034 | 0.098 |

### 4.4.1 Discussions

It can be seen from the result tables shown above that:

1. The optimization problem converges after few generations as the fitness value saturates.

2. All the constraints are satisfied and the final minimized cost has no penalty value in it.

3. The final results are obtained from multiple executions as GA may prematurely converge sometimes.

4. Monte Carlo simulations with 400000 samples were performed for validation but GA gave better results.

### 4.5 CONCLUSIONS

In this chapter, TNEP using AC model is implemented by Genetic Algorithm. The main significance of this method is the advantages of using AC model over DC model.

TNEP using AC model is implemented on Garver's 6 bus test system. Fast Decoupled Load flow was used for the AC model.

From the results it is easily observed that AC TNEP without security constraints costs 3.11 Million $ and with security constraints costs 3.49 Million $ due to more addition of lines to satisfy the N-1 contingency criteria.

The TNEP using AC model is mainly required for integration with Reactive Power Planning as the later requires AC model. In DC model line additions takes place only to carry real power from generator buses to the load buses and for the realistic system this may not provide the optimal solution. The DC TNEP may be infeasible for the actual AC system as there are reactive power flows associated with each line which must be controlled by addition of reactive VAr sources at the load buses. Hence AC TNEP if integrated with RPP may provide better plans which are looked upon in the next chapter.

# CHAPTER 5

# COMPOSITE TRANSMISSION NETWORK EXPANSION PLANNING (TNEP) AND REACTIVE POWER EXPANSION PLANNING (RPP)

## 5.1    INTRODUCTION

Transmission network expansion planning (TNEP) is a crucial issue especially in restructured power systems. In the new environment of the electricity industry, open access to transmission networks introduces some new challenges to all market participants. As electricity consumption grows rapidly, additional transmission lines are required to facilitate alternative paths for power transfer from power plants to load centres.

TNEP via simplified models such as the transportation model, hybrid model, linear disjunctive model, and DC model, among others, will usually fail to support a solution that can handle real network requirements. In this work transmission planning using an AC model should is associated with Reactive Power Planning (RPP). Without considering reactive sources or VAr-plants, the AC-TNEP problem may have an optimum solution in which only the generators satisfy reactive load demands. Although in case of generator capability of supporting reactive load demands, transferring such an amount of reactive power may reduce the available transfer capability (ATC) that lead to more new transmission lines. While by allocating VAr-plants close to the load centres, this may be possible to prevent the necessity of additional transmission lines. On the other hand, without considering the VAr-plants, load bus voltages may differ from their specified magnitudes, which may not only cause unacceptable power quality but also increase real power losses. Increasing power losses may require more transmission line additions.

Besides increasing the capacity of transmission lines and power loss reduction due to inclusion of RPP in TNEP, voltage profile enhancement and voltage stability margin improvement can be attained. The objective function of composite TNEP RPP is to determine where, how many and when new equipment such as transmission lines and reactive power sources must be added to the network in order to make its operation viable for a pre-defined planning horizon at minimum cost.

It should be noted that there are a few reports in existing planning methods discussing both stages in an integrated way. A combinatorial mathematical model in tandem with a metaheuristic technique for solving transmission network expansion planning (TNEP) using an AC model associated with reactive power planning (RPP) is presented Rahmania, Rashidinejada, Carrenoc, Romerob in [37].

In this chapter, a novel technique for integration of TNEP and RPP is implemented on Garver's 6 bus test system. The following flow chart describes the basic methodology for carrying out the decomposed integration process of TNEP and RPP.

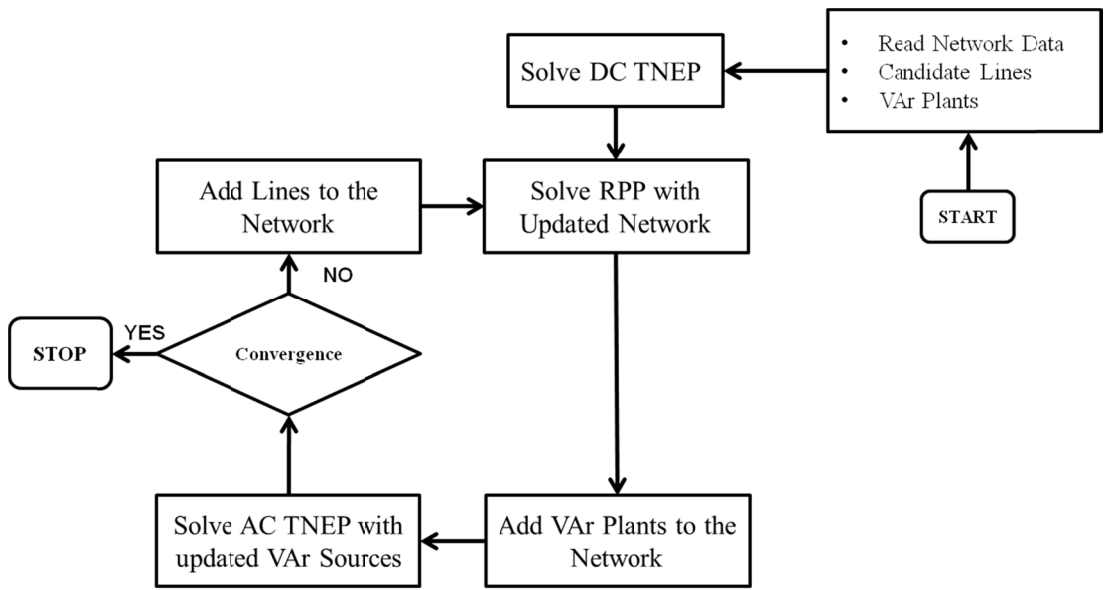

**Fig 5.1 Flow Chart for Composite TNEP RPP**

Fig. 5.1 describes the algorithm for integrated TNEP and RPP. The algorithm starts with solving a TNEP using DC model to provide an initial guess to the algorithm. Next RPP is solved with the updated network topology followed by AC TNEP. Now with this new network topology RPP is performed once again to improve its previous solution and this iterative process between the RPP and AC TNEP is continued until there is no further improvement total cost of TNEP and RPP.

## 5.2    PROBLEM FORMULATION

In this section the mathematical modeling for RPP and AC TNEP are described in details.

### 5.2.1    Mathematical Modeling for Reactive Power Planning (RPP)

The RPP problem, stated as an optimization problem, where the cost of the installation of new reactive power sources considering the fixed or installment costs as well as the variable costs and the cost of the active transmission power losses are minimized, subject to the following constraints that define satisfactory operation.

1.   Network energy balance equality constraints;

2.   Limits on: reactive power generations, bus voltage magnitudes.

These constraints must be considered for each of the relevant system scenarios, such as peak load and minimum load, at the relevant network configurations. In this problem only capacitors are considered as reactive power sources.

The RPP problem can be stated as follows

$$\min f = \sum_{i=1}^{N_{sh}} \left( d_{ci} \times Y_{ci} + k_{ci} \times q_i \right) + k_1 \times P_{Loss} \tag{5.1}$$

Subjected to

$$P_i = V_i^2 G_{ii} + \sum_{i=1}^{N_B} (V_i V_j Y_{ij} \cos(\theta_{ij} - \delta_i - \delta_j)) \tag{5.2}$$

$$Q_i = V_i^2 B_{ii} + \sum_{i=1}^{N_B} (V_i V_j Y_{ij} \sin(\theta_{ij} - \delta_i - \delta_j)) \qquad \text{i=1,2…………}N_{BUS} \tag{5.3}$$

Voltage limits

$$V_{i\min} \leq V_i \leq V_{i\max} \qquad \text{i=1,2....N}_{\text{BUS}} \qquad (5.4)$$

VAR générations limits

$$Q_{i\min} \leq Q_i \leq Q_{i\max} \qquad \text{i=1,2....N}_{\text{PV}} \qquad (5.5)$$

VAR installation limits

$$q_{i\min} \leq q_i \leq q_{i\max} \qquad \text{i=1,2......N}_{\text{SH}} \qquad (5.6)$$

$d_{ci}$ : Binary variable that decides the location VAR source.

$Y_{ci}$ : Fixed cost associated with the installation of reactive power source

$k_{ci}$ : Variable cost associated with the reactive power sources

$k_l$ : Conversion coefficient from loss to cost

$P_{Loss}$ : Total Power loss in the network

q: Shunt size

n: Number of buses

### 5.2.2   Mathematical Modeling for Transmission Network expansion Planning using AC model (AC TNEP)

$$Min\ v = c^T n \qquad (5.1)$$

Subject to

$$P(v, \theta, n) - P_G + P_D = 0 \qquad (5.2)$$

$$Q(v, \theta, n) - Q_G + Q_D = 0 \qquad (5.3)$$

$$\underline{P_G} \leq P_G \leq \overline{P_G} \qquad (5.4)$$

$$\underline{Q_G} \leq Q_G \leq \overline{Q_G} \qquad (5.5)$$

$$\underline{V} \leq V \leq \overline{V} \qquad (5.6)$$

$$(N + N_0)S^{from} \leq (N + N_0)\overline{S} \qquad (5.7)$$

$$(N + N_0)S^{to} \leq (N + N_0)\overline{S} \qquad (5.8)$$

$$0 \leq n \leq \underline{n} \qquad (5.9)$$

$n$ integer and $\theta$ unbounded

Where,

$c$ : circuit cost vector that can be added to the network

$n$ : added circuit vector, respectively.

$N, N_0$ : diagonal matrices containing vector $n$ and the existing circuits in the base configuration, respectively.

$v$ : investment due to the addition of circuits to the networks.

$\overline{n}$ : vector containing the maximum number of circuits that can be added.

$\theta$ : phase angle vector.

$P_G$ , $Q_G$ : real and reactive power generation vectors.

$P_D$, $Q_D$ : real and reactive power demand vectors;

$V$ is the voltage magnitude vector;

$\overline{P_G}, \overline{Q_G}, \overline{V}$ : vectors of maximum limits of generation of real power, reactive power and voltage magnitudes,

$\underline{P_G}, \underline{Q_G}, \underline{V}$, : vectors of minimum limits of generation of real power, reactive power and voltage magnitudes,

$S^{from}$, $S^{to}$, $\overline{S}$ : apparent power flow vectors (MVA) in the branches in both terminals and their limits.

The elements of vector $P(v, \theta, n)$ and $Q(v, \theta, n)$ are calculated by

$$P_i(v, \theta, n) = V_i \sum_{j \in N} V_j [G_{ij}(n) \cos \theta_{ij} + B_{ij}(n) \sin \theta_{ij}] \qquad (5.10)$$

$$Q_i(v, \theta, n) = V_i \sum_{j \in N} V_j [G_{ij}(n) \sin \theta_{ij} - B_{ij}(n) \cos \theta_{ij}] \qquad (5.11)$$

where i, $j \in N$ represent the buses and N is the set of all buses, $ij$ represents the circuit between buses $i$ and $j$. The bus admittance matrix elements are

$$G = \begin{cases} G_{ij}(n) = -(n_{ij}g_{ij} + n^0{}_{ij}g^0{}_{ij}) \\ G_{ii}(n) = \sum_{j \in \Omega_i}(n_{ij}g_{ij} + n^0{}_{ij}g^0{}_{ij}) \end{cases}$$

(5.12)

$$B = \begin{cases} B_{ij}(n) = -(n_{ij}b_{ij} + n^0{}_{ij}b^0{}_{ij}) \\ B_{ii}(n) = b^{sh}{}_i + \sum_{j \in \Omega_i} \begin{matrix}[n_{ij}(b_{ij} + b^{sh}{}_{ij}) \\ + n^0{}_{ij}(b^0{}_{ij} + (b^{sh}{}_{ij})^0)]\end{matrix} \end{cases}$$

where $\Omega_i$ represents the set of all buses directly connected to bus $i$; $g_{ij}$, $b_{ij}$ and $b_{ij}{}^{sh}$ are the conductance, susceptance and shunt susceptance of the transmission line or transformerij (if ij is a transformer $b_{ij}{}^{sh} = 0$), respectively, and $b_i{}^{sh}$ is the shunt susceptance at bus $i$.

Elements (ij) of vectors S ($S^{from}$, $S^{to}$) are given by

$$S_{ij} = \sqrt{P_{ij}{}^2 + Q_{ij}{}^2} \qquad (4.13)$$

$$P_{ij} = V_j^2 g_{ij} - V_i V_j(g_{ij}\cos\theta_{ij} - b_{ij}\sin\theta_{ij}) \qquad (4.14)$$

$$Q_{ij} = -V_j^2(b_{ij} + b^{sh}{}_{ij}) + V_i V_j(g_{ij}\sin\theta_{ij} + b_{ij}\cos\theta_{ij}) \quad (4.15)$$

Security Constraints is an important part of TNEP. Here, **N-1** contingency analysis is used to check the system security. The **N-1** contingency analysis looks at the system state after a single line outage. Results for both with and without Security Constraints are carried out.

## 5.3    SOLUTION METHODOLOGY

**For RPP**

Particle Swarm Optimization (PSO) is used for solving RPP

**Variables**:    VAr Sources (Size and location)

PV bus voltage setting

Tap Settings

Conversion coefficient from loss to cost : 0.06$/KWH

Fixed cost associated with the installation of reactive power source : 1000$

Variable cost associated with the reactive power sources : 30$/KVAR

$q_{min}$:Minimum size of the Capacitor  0

$q_{max}$:Maximum size of the Capacitor  48 MVAR

### Table 5.1 Particle swarm Optimization Parameters

| PSO Parameters | Value |
|---|---|
| **Population Size** | 80 |
| **Maximum Generations / Iterations** | 200 |
| $w_{max}$ $w_{min}$ $c_1$ $c_2$ | 0.9 0.3 2.1 2.1 |

**For TNEP using AC Model**

Simple Genetic Algorithm is used for solving the TNEP using AC model.

For Garver's 6bus system :

Bits required for each Transmission line to be added is 4 as there are 15 possible lines for Garver's 6 bus system  ($2^4$ =16)

Therefore, for 15 lines that can be added at max, string length becomes 15×4=60

**Fitness Function:**

$$fitness = \frac{\alpha}{1+J} \tag{2.21}$$

Where α is a constant, J is the sum of the objective function and penalties for any constraint violation

**Parent Selection Method:** Tournament Selection

**Crossover Technique:** Uniform Crossover

<div align="center">

**Table 5.2 Genetic Algorithm Parameters**

</div>

| GA Parameters | Value |
|---|---|
| Population Size | 80 |
| Maximum Generations / Iterations | 200 |
| Probabilities of Crossover and Mutation | 0.9, 0.01 |
| Number of Elite Strings | 5 |

For the implementation of Integrated TNEP RPP, load is assumed to have a constant power factor of 0.9 and the optimization is done for three load scenarios, viz. 22.5% above base case for duration of 876 hrs, base case for 6132 hrs and 30 % below base case for 1752 hrs for a yearly load duration curve. Fast decoupled Load Flow (FDLF) is used for evaluation of the state variables and hence the injections, flows and losses.

<div align="center">

**Table 5.3 Load Scenarios**

</div>

| Load | Duration (Hrs) |
|---|---|
| 22.5% above base case | 876 |
| Base Case | 6132 |
| 30% below base case | 1752 |

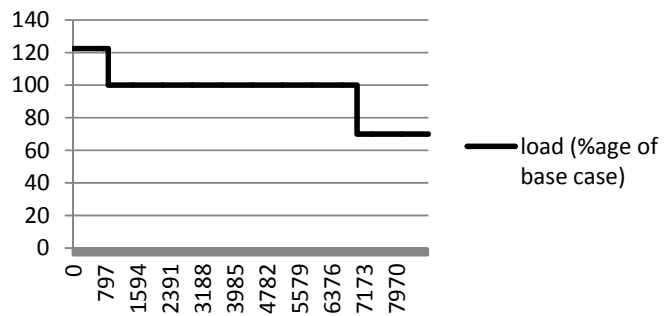

**5.4 RESULTS**

The Integrated AC TNEP and RPP were implemented on Garver's 6 bus system.

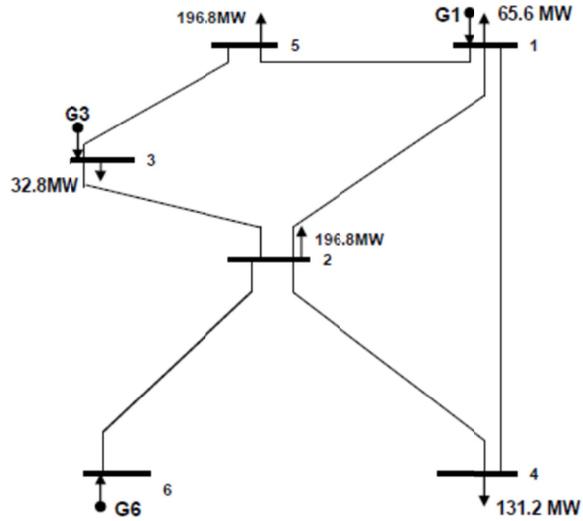

**Fig 5.2 Garver's 6 bus system at base before AC TNEP**

**Table 5.4 Technical and Economic Data of the candidate lines**

| Line # | From Bus | To Bus | Capacity (MVA) | Cost(million $) |
|--------|----------|--------|----------------|-----------------|
| 1 | 1 | 2 | 120 | 40 |
| 2 | 1 | 3 | 120 | 38 |
| 3 | 1 | 4 | 100 | 60 |
| 4 | 1 | 5 | 120 | 20 |
| 5 | 1 | 6 | 90 | 68 |
| 6 | 2 | 3 | 120 | 20 |
| 7 | 2 | 4 | 120 | 40 |
| 8 | 2 | 5 | 120 | 31 |
| 9 | 6 | 2 | 120 | 30 |
| 10 | 3 | 4 | 120 | 59 |
| 11 | 3 | 5 | 120 | 20 |
| 12 | 6 | 3 | 120 | 48 |
| 13 | 4 | 5 | 95 | 63 |
| 14 | 4 | 6 | 120 | 30 |
| 15 | 5 | 6 | 98 | 61 |

**Results for Integrated AC TNEP-RPP**

**(without Security Constraints)**

**Minimized Cost for AC TNEP : 2.2×10$^8$ $**

**Minimized Cost for RPP : 1.2782×10$^6$ $**

**Total VAr installation cost : 0.903×10$^6$ $**

**Installed VAr :**

**9 MVAR at bus 2**

**14 MVAR at bus 4**

**7 MVAR at bus 5**

**Total Cost of composite ACTNEP-RPP = 2.212782×10$^8$ $**

Table 5.5 Addition of lines for the given load scenarios and pf 0.9 for

Garver's 6 bus  system without security constraints (Integrated TNEP RPP)

| Line # | From Bus | To Bus | Lines Present | Lines Added |
|--------|----------|--------|---------------|-------------|
| 1 | 1 | 2 | 1 | 0 |
| 2 | 1 | 3 | 0 | 0 |
| 3 | 1 | 4 | 1 | 0 |
| 4 | 1 | 5 | 1 | 0 |
| 5 | 1 | 6 | 0 | 0 |
| 6 | 2 | 3 | 1 | 0 |
| 7 | 2 | 4 | 1 | 0 |
| 8 | 2 | 5 | 0 | 0 |
| 9 | 6 | 2 | 1 | **4** |
| 10 | 3 | 4 | 0 | 0 |
| 11 | 3 | 5 | 1 | **2** |
| 12 | 6 | 3 | 0 | 0 |
| 13 | 4 | 5 | 0 | 0 |
| 14 | 4 | 6 | 0 | **2** |
| 15 | 5 | 6 | 0 | 0 |

| Results for AC TNEP without Security Constraints  followed by RPP (Separate Approach) |
|---|
| Minimized Cost for AC TNEP : $3.11 \times 10^8$ \$ |

Minimized Cost for RPP :  $0.896174 \times 10^6$ \$

Total VAr installation cost : $0.543 \times 10^6$ \$

Installed VAr :

8 MVAR at bus 2

5 MVAR at bus 4

5 MVAR at bus 5

Total Cost of ACTNEP and RPP = $3.11896 \times 10^8$ \$

Table 5.6 Addition of lines for the given load scenarios and pf 0.9 for Garver's 6 bus system by AC TNEP without Security Constraints (Separate TNEP and RPP)

| Line # | From Bus | To Bus | Lines Present | Lines Added |
|---|---|---|---|---|
| 1 | 1 | 2 | 1 | 0 |
| 2 | 1 | 3 | 0 | 0 |
| 3 | 1 | 4 | 1 | 0 |
| 4 | 1 | 5 | 1 | 0 |
| 5 | 1 | 6 | 0 | 0 |
| 6 | 2 | 3 | 1 | 0 |
| 7 | 2 | 4 | 1 | 0 |
| 8 | 2 | 5 | 0 | 0 |
| 9 | 6 | 2 | 1 | 4 |
| 10 | 3 | 4 | 0 | 0 |
| 11 | 3 | 5 | 1 | 2 |
| 12 | 6 | 3 | 0 | 0 |
| 13 | 4 | 5 | 0 | 0 |
| 14 | 4 | 6 | 0 | 3 |
| 15 | 5 | 6 | 0 | 1 |

| Results for Integrated AC TNEP-RPP |
|---|
| *(with Security Constraints)* |

**Minimized Cost for AC TNEP :  $3.0 \times 10^8$ \$**

**Minimized Cost for RPP :  $1.1609 \times 10^6$ \$**

**Total VAr installation cost : $0.723 \times 10^6$ \$**

**Installed VAr :**

**11 MVAR at bus 2**

**13 MVAR at bus 4**

**3 MVAR at bus 5**

**Total Cost of composite TNEP-RPP = $3.01169 \times 10^8$ \$**

**Table 5.7 Addition of lines for the given load scenarios and pf 0.9 for Garver's 6 bus system with security constraints (Integrated TNEP RPP)**

| Line # | From Bus | To Bus | Lines Present | Lines Added |
|---|---|---|---|---|
| 1 | 1 | 2 | 1 | 0 |
| 2 | 1 | 3 | 0 | 0 |
| 3 | 1 | 4 | 1 | 0 |
| 4 | 1 | 5 | 1 | 1 |
| 5 | 1 | 6 | 0 | 0 |
| 6 | 2 | 3 | 1 | 0 |
| 7 | 2 | 4 | 1 | 0 |
| 8 | 2 | 5 | 0 | 0 |
| 9 | 6 | 2 | 1 | 4 |
| 10 | 3 | 4 | 0 | 0 |
| 11 | 3 | 5 | 1 | 5 |
| 12 | 6 | 3 | 0 | 0 |
| 13 | 4 | 5 | 0 | 0 |
| 14 | 4 | 6 | 0 | 2 |
| 15 | 5 | 6 | 0 | 0 |

**_Results for AC TNEP with Security Constraints  followed by RPP_**

**_(Separate Approach)_**

**Minimized Cost for AC TNEP :  $3.49 \times 10^8$ \$**

**Minimized Cost for RPP :  $0.87829 \times 10^6$ \$**

**Total VAr installation cost : $0.452 \times 10^6$ \$**

**Installed VAr :**

**6 MVAR at bus 2**

**11 MVAR at bus 4**

**Total Cost of ACTNEP and RPP = $3.4988 \times 10^8$ \$**

**Table 5.8 Addition of lines for the given load scenarios and pf 0.9 for Garver's 6 bus system by AC TNEP with Security Constraints (Separate TNEP and RPP)**

| Line # | From Bus | To Bus | Lines Present | Lines Added |
|--------|----------|--------|---------------|-------------|
| 1 | 1 | 2 | 1 | 0 |
| 2 | 1 | 3 | 0 | 1 |
| 3 | 1 | 4 | 1 | 0 |
| 4 | 1 | 5 | 1 | 0 |
| 5 | 1 | 6 | 0 | 0 |
| 6 | 2 | 3 | 1 | 0 |
| 7 | 2 | 4 | 1 | 0 |
| 8 | 2 | 5 | 0 | 0 |
| 9 | 6 | 2 | 1 | 4 |
| 10 | 3 | 4 | 0 | 0 |
| 11 | 3 | 5 | 1 | 2 |
| 12 | 6 | 3 | 0 | 0 |
| 13 | 4 | 5 | 0 | 0 |
| 14 | 4 | 6 | 0 | 3 |
| 15 | 5 | 6 | 0 | 1 |

### 5.4.1 Discussions

It can be seen from the result tables shown above that:

1. We can observe that the composite TNEP RPP the iterative solution decreases the total cost of TNEP and RPP and finally converges as no further improvement is possible.

2. The optimization problem converges after few generations as the fitness value saturates.

3. All the constraints are satisfied and the final minimized cost has no penalty value in it.

### 5.5 CONCLUSIONS

In this chapter, a novel approach to integrate AC TNEP and RPP is implemented. The main significance of this chapter is the decomposed approach in which TNEP and RPP are performed.

The integrated TNEP and RPP was implemented on Garver's bus test system.

From the results it can be observed that AC TNEP without RPP, implemented in previous chapter was adding more lines for transfer of reactive power.

In the integrated approach implemented here, the RPP added reactive power sources at the load buses which made the AC TNEP to add fewer lines thus reducing the total cost of investment. The total cost for composite TNEP RPP is economical than separate TNEP and RPP by 29% without security constraints.

In the previous chapter TNEP using AC model was performed. The cost of plan was high as there was no reactive power source and all reactive power had to be supplied by the generators. Hence, extra lines were required to transfer this power. But with this integrated TNEP RPP approach it is possible to add reactive sources at the load end minimizing the transmission lines to be added. Moreover, using the traditional DC TNEP and then RPP separately may involve more investment in adding reactive power sources as the lines added in DC TNEP only takes care of the real power flow.

# CHAPTER 6

# TRANSMISSION NETWORK EXPANSION PLANNING USING INTERIOR-POINT METHOD

## 6.1    INTRODUCTION

The TNEP is a very complex problem that deals with continuous and integer variables and is classified as a nonlinear integer mixed programming problem that is difficult to solve, especially for large-scale power systems. Traditionally, these problems are handled by heuristics [35], combinatorial algorithms and recently by metaheuristic algorithms[29-33]. Recently, some authors have used Interior Point Method (IP) to tackle TNEP problem [54].

The major technique involved in the modelling of TNEP using IP methods is the handling of the integer variables as IP methods are capable of handling continuous variables only. The authors in [54] have proposed the use of sigmoid functions to convert continuous variables to discrete variables.

The DC model is modified to include the power losses in the TNEP model. In this approach, the active power flow ($P_{ij}$) in a given line $i$-$j$ is expressed by

$$P_{ij} = b_{ij}\theta_{ij} + g_{ij}\frac{\theta_{ij}^2}{2}, \qquad \forall (i,j) \in E \qquad (6.1)$$

where

$b_{ij}$ is the susceptance of branch $i$-$j$ ,

$\theta_{ij}$ is the angular difference between busbars  $i$-$j$ ($\theta_{ij} = \theta_i - \theta_j$) ,

$g_{ij}$ is the conductance of branch $i$-$j$ , $E$ and is the set of existing branches of the network.

The decision of whether or not to include a candidate transmission line to be constructed requires the modification of (6.1) leading to (6.2)

$$P_{ij} = ED_{ij}\left[b_{ij}\theta_{ij} + g_{ij}\frac{\theta_{ij}^2}{2}\right], \qquad \forall (i,j) \in C \qquad (6.2)$$

where $ED_{ij}$ is the expansion decision function given by (6.3), and $C$ is the set of transmission lines that are candidates to expansion

$$ED_{ij}(x_{ij}) = \frac{e^{\alpha u_{ij}} - 1}{e^{\alpha u_{ij}} + 1} \tag{6.3}$$

where $\alpha$ is the sigmoid slope (adopted $\alpha$ =1) and $u_{ij}$ is the decision variable.

The argument of the expansion decision function ranges from 0 to 20, as shown in Fig. 6.1. The upper-bound argument of the sigmoid function can be any number that makes as close as possible to 1.

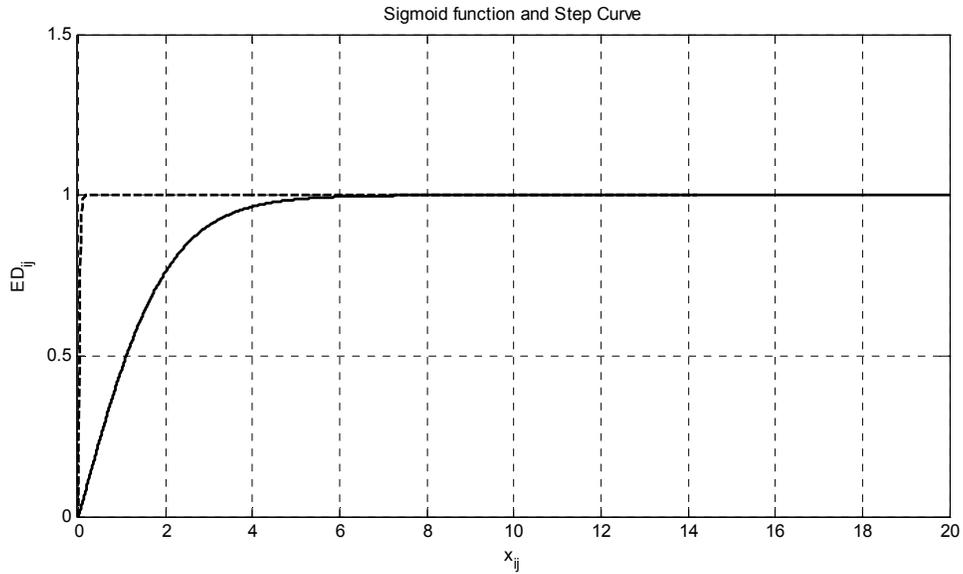

Fig 6.1 Sigmoid Function ($ED_{ij}$)

## 6.1    PROBLEM FORMULATION

The OPF problem can be formulated as

$$Min \sum_{(i,j) \in C} c_{ij} ED_{ij}(u_{ij}) \tag{6.4}$$

subject to

$$G_i - \sum_{(i,j) \in E} P_{ij} - \sum_{(i,j) \in C} P_{ij} - d_i = 0, \quad i = 1 \text{ to } N_{bus} \tag{6.5}$$

$$-P_{ij}^{max} \leq P_{ij} \leq P_{ij}^{max}, \qquad \forall (i,j) \in E, C \tag{6.6}$$

$$0 \leq u_{ij} \leq 20, \qquad \forall (i,j) \in C \tag{6.7}$$

where

$c_{ij}$ is expansion cost of the transmission line $i - j$ (US$);

$u_{ij}$ is argument of the sigmoid function;

$G_i$ is generation unit at busbar ;

$d_i$ is forecasted load demand at busbar ;

$P_{ij}$ active power flow of branch $i$ - $j$ . Existing branchesare given by (6.1).Candidate transmission lines are given by (6.2);

$P_{ij}{}^{max}$ active power flow limit of transmission line $i$ - $j$ .

The objective function (6.4) represents power system investment costs. Equation (6.5) corresponds to the expressions of the dc power flow. The constraints (6.6) are related to active power flow limits on the transmission lines. The expression (6.7) represents the lower and upper limits of the optimization variables.

## 6.2    SOLUTION METHODOLOGY

Here the optimization variables are $u_{ij}$ and $\theta_i$ (i $\in$ 2 $to$ $N_{bus}$) assuming 1 to be reference bus. For a particular corridor $u_{ij}$ consists of as many variables as the maximum no. of lines possible in that corridor say $k_{max}$. The solution vector is of the form

$$x = \begin{bmatrix} \theta_2 \\ \vdots \\ \theta_i \\ \vdots \\ \theta_{N_{bus}} \\ \vdots \\ u_{ij}{}^1 \\ \vdots \\ u_{ij}{}^k \\ \vdots \\ u_{ij}{}^{kmax} \\ \vdots \end{bmatrix} \qquad (6.8)$$

The problem can be written as
$$Min\, f(x)$$
subject to

$$g(x) = 0$$

$$h^{min} \leq h(x) \leq h^{max} \qquad (6.9)$$

$$x^{min} \leq x \leq x^{max}$$

Transforming inequalities into equality constraints

$$Min\ f(x)$$

subject to

$$g(x) = 0$$

$$-s_1 - s_2 + h^{max} - h^{min} = 0$$

$$-h(x) - s_2 + h^{max} = 0 \qquad (6.10)$$

$$-s_3 - s_4 + x^{max} - x^{min} = 0$$

$$-x - s_4 + x^{max} = 0$$

$$s_1, s_2, s_3, s_4 \geq 0$$

Non negativity of slack variables are handled by incorporating them in logarithmic barrier terms

$$Min\ f(x) - \mu^k \sum_{l=1}^{p}(ln(s_1)^l + ln(s_2)^l) - \mu^k \sum_{l=1}^{q}(ln(s_3)^l + ln(s_4)^l)$$

subject to

$$g(x) = 0$$

$$-s_1 - s_2 + h^{max} - h^{min} = 0$$

$$-h(x) - s_2 + h^{max} = 0 \qquad (6.11)$$

$$-s_3 - s_4 + x^{max} - x^{min} = 0$$

$$-x - s_4 + x^{max} = 0$$

$$s_1, s_2, s_3, s_4 > 0$$

Lagrangian can be written as

$$L_\mu = f(x) - \mu^k \sum_{i=1}^{p}(lns_{1_i} + lns_{2_i}) - \mu^k \sum_{j=1}^{q}(lns_{3_j} + lns_{4_j}) - \lambda^T g(x)$$

$$+ z_1^T(h^{max} - h^{min} + s_1 + s_2) + z_2^T(h(x) - h^{max} + s_2)$$

$$+ z_3^T(x^{max} - x^{min} + s_3 + s_4) + z_4^T(x - x^{max} + s_4)$$

$$(6.12)$$

where $\lambda$, $z_1$, $z_2$, $z_3$ and $z_4$ are vectors of Lagrangian multipliers, called the dual variables. $x$, $s_1$, $s_2$, $s_3$ and $s_4$ are vectors known as the primal variables.

The local minimiser $x^*$ for each $\mu^k$ must satisfy the Karush-Kuhn-Tucker (KKT) conditions. These KKT conditions are the first order derivatives of the Lagrangian function (6.12) with respect to all variables present in the function. Thus,

$$\nabla_x L_\mu = 0 = \nabla f(x) - J_g^T(x)\lambda + J_h^T(x)z_2 + z_4$$

$$\nabla_{s_{1_i}} L_\mu = 0 = -\mu^k s_{1_i}^{-1} + z_{1_i} \text{ (In matrix form; } -\mu^k S_1^{-1}a + z_1 = 0)$$

$$\nabla_{s_{2_i}} L_\mu = 0 = -\mu^k s_{2_i}^{-1} + z_{1_i} + z_{2_i} \text{ (In matrix form; } -\mu^k S_2^{-1}a + z_1 + z_2 = 0)$$

$$\nabla_{s_{3_i}} L_\mu = 0 = -\mu^k s_{3_i}^{-1} + z_{3_i} \text{ (In matrix form; } -\mu^k S_3^{-1}a + z_3 = 0)$$

$$\nabla_{s_{4_i}} L_\mu = 0 = -\mu^k s_{4_i}^{-1} + z_{3_i} + z_{4_i} \text{ (In matrix form; } -\mu^k S_4^{-1}a + z_3 + z_4 = 0)$$

$$\nabla_\lambda L_\mu = 0 = -g(x)$$

$$\nabla_{z_1} L_\mu = 0 = h^{max} - h^{min} + s_1 + s_2$$

$$\nabla_{z_2} L_\mu = 0 = h(x) - h^{max} + s_2$$

$$\nabla_{z_3} L_\mu = 0 = x^{max} - x^{min} + s_3 + s_4$$

$$\nabla_{z_4} L_\mu = 0 = x - x^{max} + s_4$$

$$(6.13)$$

where $\nabla$ is the gradient of $f(x)$ with respect to $x$, $J_g$ is Jacobian of $g(x)$ and $J_h$ is the Jacobian of $h(x)$. $S_1$, $S_2$, $S_3$ and $S_4$ are diagonal matrices defined by the diagonal components as $s_1$, $s_2$, $s_3$, and $s_4$ respectively and $a$ stands for vectors of ones of appropriate dimension.

These KKT conditions are non-linear equations. Their solution however is approximated by one iteration of Newton`s method. The Newton`s direction is only a means to follow the path of minimizers parameterized by $\mu^k$. So for obtaining the solution, we linearize the KKT equations (6.13) about some sub-optimal point defined by $(x_0, \lambda_0, s_{1_0}, s_{2_0}, s_{3_0}, s_{4_0}, z_{1_0}, z_{2_0}, z_{3_0}, z_{4_0})$;

$\nabla f(x_0) + \nabla^2 f(x_0)\Delta x - J_g^T(x_0)\Delta\lambda - J_g^T(x_0)\lambda_0 - \lambda_0\nabla J_g^T(x_0)\Delta x +$

$J_h^T(x_0)\Delta z_2 + J_h^T(x_0)z_{20} + z_{20}\nabla J_h^T(x_0)\Delta x + z_{40} + \Delta z_4 = 0$      (6.14)

$-\mu^k a + S_1^0 Z_1^0 + S_1^0\Delta Z_1 + \Delta S_1 Z_1^0 = 0$      (6.15)

$-\mu^k a + S_2^0(Z_1^0 + Z_2^0) + S_2^0(\Delta Z_1 + \Delta Z_2) + \Delta S_2(Z_1^0 + Z_2^0) = 0$      (6.16)

$-\mu^k a + S_3^0 Z_3^0 + S_3^0\Delta Z_3 + \Delta S_3 Z_3^0 = 0$      (6.17)

$-\mu^k a + S_4^0(Z_3^0 + Z_4^0) + S_4^0(\Delta Z_3 + \Delta Z_4) + \Delta S_4(Z_3^0 + Z_4^0) = 0$      (6.18)

$g(x_0) + J_g(x_0)\Delta x = 0$      (6.19)

$h^{max} - h^{min} + S_1^0 a + S_2^0 a + \Delta s_1 + \Delta s_2 = 0$      (6.20)

$h(x_0) + J_h(x_0)\Delta x - h^{max} + S_2^0 a + \Delta s_2 = 0$      (6.21)

$x^{max} - x^{min} + S_3^0 a + S_4^0 a + \Delta s_3 + \Delta s_4 = 0$      (6.22)

$x_0 + \Delta x - x^{max} + S_4^0 a + \Delta s_4 = 0$      (6.23)

Equations (6.15)-(6.18) represent the complimentary conditions. Rearranging these equations (6.14)-(6.23) in matrix form, we get:

$$\begin{bmatrix} \nabla_x^2 L_\mu & -J_g^T(x_0) & 0 & 0 & 0 & J_h^T(x_0) & 0 & 0 & 0 & I \\ 0 & 0 & Z_1^0 & S_1^0 & 0 & 0 & 0 & 0 & 0 & 0 \\ 0 & 0 & 0 & S_2^0 & (Z_1^0+Z_2^0) & S_2^0 & 0 & 0 & 0 & 0 \\ 0 & 0 & 0 & 0 & 0 & 0 & Z_3^0 & S_3^0 & 0 & 0 \\ 0 & 0 & 0 & 0 & 0 & 0 & 0 & S_4^0 & (Z_3^0+Z_4^0) & S_4^0 \\ J_g(x_0) & 0 & 0 & 0 & 0 & 0 & 0 & 0 & 0 & 0 \\ 0 & 0 & I & 0 & I & 0 & 0 & 0 & 0 & 0 \\ J_h(x_0) & 0 & 0 & 0 & I & 0 & 0 & 0 & 0 & 0 \\ 0 & 0 & 0 & 0 & 0 & 0 & I & 0 & I & 0 \\ I & 0 & 0 & 0 & 0 & 0 & 0 & 0 & I & 0 \end{bmatrix} \begin{bmatrix} \Delta x \\ \Delta\lambda \\ \Delta s_1 \\ \Delta z_1 \\ \Delta s_2 \\ \Delta z_2 \\ \Delta s_3 \\ \Delta z_3 \\ \Delta s_4 \\ \Delta z_4 \end{bmatrix} = \begin{bmatrix} r_x \\ r_2 \\ r_3 \\ r_4 \\ r_5 \\ r_6 \\ r_7 \\ r_8 \\ r_9 \\ r_{10} \end{bmatrix}$$

(6.24)

where $\nabla_x^2 L_\mu = \nabla^2 f(x_0) - \lambda_0\nabla J_g^T(x_0) + z_{20}\nabla J_h^T(x_0)$

$r_x = -\nabla f(x_0) + J_g^T(x_0)\lambda_0 - J_h^T(x_0)z_{20} - z_{40}$

$r_2 = \mu^k a - S_1^0 Z_1^0$

$r_3 = \mu^k a - S_2^0(Z_1^0 + Z_2^0)$

$r_4 = \mu^k a - S_3^0 Z_3^0$

$r_5 = \mu^k a - S_4^0(Z_3^0 + Z_4^0)$

$r_6 = -g(x_0)$

$$r_7 = h^{max} - h^{min} - S_1{}^0 - S_2{}^0$$

$$r_8 = h^{max} - S_2{}^0 - h(x_0)$$

$$r_9 = x^{max} - x^{min} - S_3{}^0 - S_4{}^0$$

$$r_{10} = x^{max} - S_4{}^0 - x_0$$

Reducing the matrix in equation (6.24) by Gauss Elimination Method, we get:

$$\begin{bmatrix} A & -J_g^T(x_0) \\ J_g(x_0) & 0 \end{bmatrix} \begin{bmatrix} \Delta x \\ \Delta \lambda \end{bmatrix} = \begin{bmatrix} r_x - J_h^T(x_0)(r_b - r_a) - (r_d - r_c) \\ r_6 \end{bmatrix} \tag{6.25}$$

where $A = \nabla_x^2 L_\mu + J_h^T(x_0)\left(S_1^{-1}Z_1 + S_2^{-1}(Z_1 + Z_2)\right)J_h(x_0) + S_3^{-1}Z_3 + S_4^{-1}(Z_3 + Z_4)$

$$r_a = S_1^{-1}(r_2 + Z_1 r_8)$$

$$r_b = S_2^{-1}(r_3 - (Z_1 + Z_2)r_8)$$

$$r_c = S_3^{-1}(r_4 + Z_3 r_{10})$$

$$r_d = S_4^{-1}(r_5 - (Z_3 + Z_4)r_{10})$$

The step changes in other dual and primal variables in terms of $\Delta x$ are:

$$\Delta s_1 = -r_8 + J_h(x_0)\Delta x \tag{6.26}$$

$$\Delta z_1 = r_a - S_1^{-1}Z_1 J_h(x_0)\Delta \tag{6.27}$$

$$\Delta s_2 = -\Delta s_2 \tag{6.28}$$

$$\Delta z_2 = r_b + S_2^{-1}(Z_1 + Z_2)J_h(x_0)\Delta x \tag{6.29}$$

$$\Delta s_3 = -r_{10} + \Delta x \tag{6.30}$$

$$\Delta z_3 = r_c - S_3^{-1}Z_3\Delta x \tag{6.31}$$

$$\Delta s_4 = -\Delta s_3 \tag{6.32}$$

$$\Delta z_4 = r_d + S_4^{-1}(Z_3 + Z_4)\Delta x - \Delta z_3 \tag{6.33}$$

Now, having obtained all the primal and dual variables and the reduced matrix form, the interior point method can be stated as follows:

1. Set $k = 0$, define $\mu^k$ and choose or guess a starting point $x$ (for power system applications generally the converged power flow solution is chosen as the starting point) and then choose other dual and primal variables as:

$$s_{1_i} = \min\{\max\left(0.25 h_i^\Delta, h_i(x) - h_l\right), 0.75 h_i^\Delta\}$$

$$s_{2_i} = h_i^\Delta - s_{1_i}$$

$$s_{3_i} = \min\{\max(0.25x_i^\Delta, x_i - x_l), 0.75x_i^\Delta\}$$

$$s_{4_i} = x_i^\Delta - s_{3_i}$$

$$Z_1 = \mu^k S_1^{-1} a$$

$$Z_2 = \mu^k S_2^{-1} a - Z_1$$

$$Z_3 = \mu^k S_3^{-1} a$$

$$Z_4 = \mu^k S_4^{-1} a - Z_3$$

$$\lambda_i = -1 \ (for \ real \ power \ equality)$$

Having obtained a starting point for all variables, solve the reduced matrix in equation (6.25) to compute the step changes in all variables in the Newton direction using equations (6.26)-(6.33).

2. Update the variables by:

$$x^{k+1} = x^k + \alpha_p^k \Delta x, \qquad s_i^{k+1} = s_i^k + \alpha_p^k \Delta s_i$$

$$\lambda^{k+1} = \lambda^k + \alpha_d^k \Delta \lambda, \qquad z_i^{k+1} = z_i^k + \alpha_d^k \Delta z_i$$

where the scalars $\alpha_p$ and $\alpha_d$ are the step length parameters of primal and dual variables respectively which $\in (0,1]$.

3. If the new point obtained satisfies the convergence criteria (equation (6.35)), terminate otherwise, set $k = k + 1$, update the barrier parameter $\mu^k$ using equation (6.34) and return to step 2.

The scalars $\alpha_p$ and $\alpha_d$ which control the step lengths of primal and dual variables respectively can be determined as:

$$\alpha_p^k = min_{i,j}\{\frac{-s_{1_i}^k}{\Delta s_{1_i}}: \Delta s_{1_i} < 0, \frac{-s_{2_i}^k}{\Delta s_{2_i}}: \Delta s_{2_i} < 0, \frac{-s_{3_j}^k}{\Delta s_{3_j}}: \Delta s_{3_j} < 0, \frac{-s_{4_j}^k}{\Delta s_{4_j}}: \Delta s_{4_j} < 0\}$$

$$\alpha_p^k = \min\{1, \gamma \alpha_p^k\}$$

$$\alpha_d^k = min_{i,j}\{\frac{-z_{1_i}^k}{\Delta z_{1_i}}: \Delta z_{1_i} < 0, \frac{-z_{3_j}^k}{\Delta z_{3_j}}: \Delta z_{3_j} < 0, \frac{-(z_{1_i}^k + z_{2_i}^k)}{\Delta z_{1_i} + \Delta z_{2_i}}: (\Delta z_{1_i} + \Delta z_{2_i})$$

$$< 0, \frac{-(z_{3_j}^k + z_{4_j}^k)}{\Delta z_{3_j} + \Delta z_{4_j}}: (\Delta z_{3_j} + \Delta z_{4_j}) < 0\}$$

$$\alpha_d^k = \min\{1, \gamma \alpha_d^k\}$$

where $\gamma \in (0,1)$ is a safety factor to ensure that the next point will satisfy the strict positively conditions; a typical value of $\gamma = 0.9995$.

Separate step lengths in the primal and dual spaces, as presented above, is an advantage of primal dual interior point methods for linear programming and has been proven highly efficient in practice, thus reducing the number of iterations to convergence by 10%-20% in a typical problem.

For general non-linear programming, the interdependence of primal and dual variables present in the dual feasibility conditions do not allow for separate step lengths in the primal and dual variables. In this case, a common step length is chosen; $\alpha_p^k = \alpha_d^k = \min\{\alpha_p^k, \alpha_d^k\}$.

For the updation of the barrier parameter $\mu^k$, the residual of the complimentary conditions, called the complimentary gap, at the current iterate is evaluated as:

$$\rho^k = \left[(z_1^k)^T s_1^k + (z_1^k + z_2^k)^T s_2^k + (z_3^k) s_3^k + (z_3^k + z_4^k)^T s_4^k\right]$$

This sequence $\{\rho^k\}_{k=0}^{\infty}$ must converge to zero. The relationship between $\mu^k$ and $\rho^k$ suggest that $\mu^k$ could be reduced based on a predicted decrease of the complimentary gap $\rho^k$, as:

$$\mu^{k+1} = \frac{\beta^k \rho^k}{2(p+q)}$$

(6.34)

where $\beta^k$ is the expected cut in the complimentary gap. The parameter $\beta^k \in (0,1)$ is called the centering parameter. If $\beta^k = 1$, the KKT conditions define a centering direction. For $\beta^k = 0$, it is affine scaling direction. To trade off between twin goals of reducing $\mu^k$ and improving centrality, $\beta^k$ is chosen as:

$$\beta^k = \max\{0.95\beta^{k-1}, 0.1\} \; with \; \beta^0 = 0.2$$

The convergence criterion is:

$$|\Delta f(x)| <\epsilon_1, |\Delta x| <\epsilon_2, \mu^k <\epsilon_\mu$$

(6.35)

where $\epsilon_1 = 0.0001, \epsilon_2 = 0.01 \, \epsilon_1$ and $\epsilon_\mu = 10^{-12}$ typically.

To implement on Garver's 6 bus system, the gradient and hessian matrices are to be created.

## 6.4    DISCUSSION

To implement on Garver's 6 bus system, the gradient and hessian matrices are to be created according to the problem.

The problem has been modeled in MATLAB but due to time constraints the final results could not be added as it has some issues regarding convergence which are yet to be sorted out.

# CHAPTER 7

# CONCLUSION AND SCOPE OF FUTURE WORK

In this thesis several problems regarding power System Resource Expansion Planning have been considered. Implementations were performed using Matlab programming. The proposed approach has been tested for standard test systems, Garver's 6 bus system and IEEE 24 bus Reliability Test system. Monte Carlo Simulations were performed for validation.

Traditionally, planning problems were mostly heuristics in nature and recently optimization techniques have been used to solve these complicated problems. Conventional methods are not usually used for solving these problems because of the discrete nature of the problems. In this thesis, some important problems of power system planning are studied.

**In chapter 2: Transmission Constrained Generation Expansion Planning (TC GEP)** is implemented in which the conventional Generation Expansion Planning is solved considering the network topology so that the final solution is feasible for the given network. Here DC load flow was performed to check for the line flows and it was observed that the GEP without transmission constraints although being economical was not feasible for the given network topology. If sufficient transmission corridor is unavailable then network expansion should be the only option. The problem was implemented in IEEE 24 bus RTS. Simple Genetic algorithm was performed for solving this problem. Both dynamic and static problems were solved.

**In chapter 3:** A novel technique for **Composite Generation and Transmission Network Expansion Planning** is implemented in which the GEP and TNEP is solved in a single optimization process. Conventionally, TNEP is performed after GEP. But this method has the disadvantage that the planning may be uneconomical although the problem is less complex. The results showed that composite GEP TNEP was more economical than separated conventional method. Here also DC load flow was performed to check for the line flows. The problem was implemented in IEEE 24 bus RTS. Simple Genetic algorithm was performed for solving this problem and the string composed of both newly added generators and transmission lines. Both dynamic and static problems were solved.

**In chapter 4: Transmission Network Expansion Planning (TNEP) using AC Model** is solved. Unlike DC model, which only takes care of real power flows, AC model gives the solution for an actual system. TNEP using AC model is required for integration with RPP. A constant power factor for the load is assumed and reactive powers are supplied by the generators. This planning is less economical than that with DC model although the solution is feasible for the actual system. The problem was implemented in Garver's 6 bus system and simple Genetic Algorithm was used. Fast decoupled load flow was performed to calculate the state variables and three different load scenarios were considered.

**In chapter 5:** A novel approach for **Composite Transmission Network Expansion Planning (TNEP) and Reactive Power Expansion Planning (RPP)** is studied and implemented. Here, TNEP using AC model is considered. This method is an iterative decomposed method in which TNEP and RPP iterate among itself so that they finally arrive at a mutual optimal solution. This method provides better solution than the conventional way of performing RPP after TNEP. The problem was implemented in Garver's 6 bus system.

**In chapter 6: Transmission Network Planning using Interior-Point Method** is studied. Here, TNEP using DC model is considered. The main issue in solving TNEP using Interior-Point (IP) methods is the handling of discrete variables. For solving this problem, sigmoid functions are used to convert the continuous variables into discrete ones. In this chapter, the DC load flow equations are modified to model the TNEP problem for IP methods. Newton's method is used to finally solve the problem. The detailed modeling and matrix structures are discussed.

**FUTURE SCOPE**

In this dissertation, the power system resource expansion planning problems are studied. The integration of GEP & TNEP and integration of TNEP and RPP are solved in this work. This can be extended to integration of all the three planning problems. Interior point techniques have been modeled for TNEP. This can be extended to model all the three planning problems. Planning problems can be extended to deregulated environment where GENCOs can optimize their plans. Planning problems can be extended to have mutiobjectives with environmental costs as one of the objectives. TNEP RPP can be performed by adding voltage stability constraints. The problems can be extended for uncertainty in load forecasts. The integrity of the algorithms can be tested for larger systems.

# *References*

[1] Jong-Bae Park; Young-Moon Park; Jong-Ryul Won; Lee, K.Y.; , "An improved genetic algorithm for generation expansion planning," *Power Systems, IEEE Transactions on* , vol.15, no.3, pp.916-922, Aug 2000

[2] Kannan, S.; Slochanal, S.M.R.; Padhy, N.P.; , "Application and comparison of metaheuristic techniques to generation expansion planning problem," *Power Systems, IEEE Transactions on* , vol.20, no.1, pp. 466- 475, Feb. 2005.

[3] Murugan, P.; Kannan, S.; Baskar, S.; , "Application of NSGA-II Algorithm to Single-Objective Transmission Constrained Generation Expansion Planning," *Power Systems, IEEE Transactions on* , vol.24, no.4, pp.1790-1797, Nov. 2009.

[4] Kannan, S.; Baskar, S.; McCalley, J.D.; Murugan, P.; , "Application of NSGA-II Algorithm to Generation Expansion Planning," *Power Systems, IEEE Transactions on* , vol.24, no.1, pp.454-461, Feb. 2009

[5] J. A. Bloom, "Long-range generation planning using decomposition and probabilistic simulation," *IEEE Trans. on PAS*, vol. 101, no. 4, pp. 797–802, 1982.

[6] Y. M. Park, K. Y. Lee, and L. T. O. Youn, "New analytical approach for long-term generation expansion planning based maximum principle and Gaussian distribution function," *IEEE Trans. on PAS*, vol. 104, pp. 390–397, 1985.

[7] A. K. David and R. Zhao, "Integrating expert systems with dynamic programming in generation expansion planning," *IEEE Trans. on PWRS*, vol. 4, no. 3, pp. 1095–1101, 1989.

[8] Y. Fukuyama and H. Chiang, "A parallel genetic algorithm for generation expansion planning," *IEEE Trans. on PWRS*, vol. 11, no. 2, pp. 955–961, 1996.

[9] D. C.Walters and G. B. Sheble, "Genetic algorithm solution of economic dispatch with valve point loading," IEEE Trans. on PWRS, vol. 8, no. 3, pp. 1325–1332, 1993.

[10] P. H. Chen and H. C. Chang, "Large-scale economic dispatch by genetic algorithm," IEEE Trans. on PWRS, vol. 10, no. 4, pp. 1919–1926, 1995.

[11] D. Dasgupta and D. R. McGregor, "Thermal unit commitment using genetic algorithms," IEE Proc.—Gener. Transm. Distrib., vol. 141, no. 5, pp. 459–465, 1994.

[12] G. B. Sheble, T. T. Maifeld, K. Brittig, and G. Fahd, "Unit commitment by genetic algorithm with penalty methods and a comparison of Lagrangian search and genetic algorithm-economic dispatch algorithm," *Int. Journal of Electric Power & Energy Systems*, vol. 18, no. 6, pp. 339–346, 1996.

[13] K. Iba, "Reactive power optimization by genetic algorithm," *IEEE Trans. on PWRS*, vol. 9, no. 2, pp. 685–692, 1994.

[14] K. Y. Lee, X. Bai, and Y. M. Park, "Optimization method for reactive power planning using a genetic algorithm," *IEEE Trans. on PWRS*, vol. 10, no. 4, pp. 1843–1850, 1995.

[15] K. Y. Lee and F. F. Yang, "Optimal reactive power planning using evolutionary algorithms: A comparative study for evolutionary programming, evolutionary strategy, genetic algorithm, and linear programming," *IEEE Trans. on PWRS*, vol. 13, no. 1, pp. 101–108, 1998.

[16] R. Dimeo and K. Y. Lee, "Boiler–Turbine control system design using a genetic algorithm," *IEEE Trans. on Energy Conversion*, vol. 10, no. 4, pp. 752–759, 1995.

[17] Guoxian Liu, Hiroshi Sasaki, Naoto Yorino, Application of network topology to long range composite expansion planning of generation and transmission lines, Electric Power Systems Research, Volume 57, Issue 3, 20 April 2001, Pages 157-162.

[18] Sepasian, M.S.; Seifi, H.; Foroud, A.A.; Hatami, A.R.; , "A Multiyear Security Constrained Hybrid Generation-Transmission Expansion Planning Algorithm Including Fuel Supply Costs," *Power Systems, IEEE Transactions on* , vol.24, no.3, pp.1609-1618, Aug. 2009

[19] A.K. David, R.D. Zhao, An expert system with fuzzy sets for optimal planning, IEEE Trans. Power Syst. 6 (1) (1991) 59–65.

[20] B. Mo, J. Hegge, I. Wamgensteen, Stochastic generation expansion planning by means of stochastic dynamic programming, IEEE Trans. Power Syst. 6 (2) (1991) 662–668.

[21] T.P. Hadjicostas, R.N. Adams, The flexible rolling schedule for infinite-horizon optimality in generation expansion planning, IEEE Trans. Power Syst. 7 (3) (1992) 1182–1188.

[22] B.G. Gorenstin, N.M. Campodonico, J.P. Costa, M.V.F. Pereira, Power system expansion planning under uncertainty, IEEE Trans. Power Syst. 8 (1) (1993) 129–136.

[23] R. Tanabe, K. Yasuda, R. Yokoyama, H. Sasaki, Flexible generation mix under multi objectives and uncertainties, IEEE Trans. Power Syst. 8 (2) (1993) 581–587.

[24] W.J. Head, H.V. Nguyen, R.L. Kahle, The procedure used to assess the long range generation and transmission resources in the mid-continent area power pool, IEEE Trans. Power Syst. 5 (4) (1990) 1137–1145.

[25] L. Wenynan, R. Billinton, A minimum cost assessment method for composite generation and transmission system expansion planning, IEEE Trans. Power Syst. 8 (2) (1993) 628–635.

[26] M.L. Baughman, S.N. Siddiqi, J.W. Zamikau, Integrating transmission into IRP — part 1: analytical approach, IEEE Trans. Power Syst. 10 (3) (1995) 1652–1659.

[27] M.L. Baughman, S.N. Siddiqi, J.W. Zamikau, Integrating transmission into IRP — part 2: case study results, IEEE Trans. Power Syst. 10 (3) (1995) 1660–1666.

[28] Q. Xia, Y.H. Song, C.Q. Kang, N.D. Xiang, Novel models and algorithms for generation unit location optimization, IEEE Trans. Power Syst. 12 (4) (1997) 1584–1590.

[29] Romero R., Gallego R.A., Monticelli A.: 'Transmission network expansion planning by simulated annealing', IEEE Trans. Power Syst., 1990, 11, (1), pp. 364–369

[30] Gallego R.A., Romero R., Monticelli A.J.: 'Tabu search algorithm for network synthesis', IEEE Trans. Power Syst., 2000, 15, (2), pp. 490–495

[31] Da Silva E.L., Gil H.A., Areiza J.M.: 'Transmission network expansion planning under an improved genetic algorithm', IEEE Trans. Power Syst., 2000, 15, (3), pp. 1168–1175

[32] Escobar A.H., Gallego R.A., Romero R.: 'Multistage and coordinated planning of the expansion of transmission systems', IEEE Trans. Power Syst., 2004, 19, (2), pp. 735–744

[33] Romero R., Rider M.J., Silva I.D.J.: 'A metaheuristic to solve the transmission expansion planning', IEEE Trans. Power Syst., 2007, 22, (4), pp. 2289–2291

[34] R. Romero, A. Monticelli, A. Garcia, and S. Haffner, "Test systems and mathematical models for transmission network expansion planning," Proc. Inst. Elect. Eng., Gen., Transm., Distrib., vol. 149, no. 1, pp. 29–36, Jan. 2002.

[35] Garver L.L.: 'Transmission network estimation using linear programming', IEEE Trans. PAS, 2007, 89, (7), pp. 1688–1697

[36] Rider, M.J.; Garcia, A.V.; Romero, R.; , "Power system transmission network expansion planning using AC model," Generation, Transmission & Distribution, IET , vol.1, no.5, pp.731-742, September 2007

[37] M. Rahmani, M. Rashidinejad, E.M. Carreno, R. Romero, Efficient method for AC transmission network expansion planning, Electric Power Systems Research, Volume 80, Issue 9, September 2010, Pages 1056-1064, ISSN 0378-7796

[38] K. Aoki, M. Fan, and A. Nishikori, "Optimal Var planning by approximation method for recursive mixed-integer linear programming," IEEE Trans. Power Syst., vol. 3, no. 4, pp. 1741–1747, Nov. 1988.

[39] J. R. S. Mantovani and A. V. Garcia, "A heuristic method for reactive power planning," IEEE Trans. Power Syst., vol. 11, no. 1, pp. 68–74, Feb. 1996.

[40] M. Delfanti, G. Granelli, P. Marannino, and M. Montagna, "Optimal capacitor placement using deterministic and genetic algorithms," IEEE Trans. Power Syst., vol. 15, no. 3, pp. 1041–1046, Aug. 2000.

[41] C. T. Hsu, Y. H. Yan, C. S. Chen, and S. L. Her, "Optimal reactive power planning for distribution systems with nonlinear loads," in Proc. IEEE Region 10 Int. Conf. Computer, Communication, Control and Power Engineering, Beijing, China, Oct. 19–21, 1993, vol. 5, pp.330–333.

[42] W. Zhang, Y. Liu, and Y. Liu, "Optimal VAr planning in area power system," in Proc. Int. Conf. Power System Technology, Oct. 13–17, 2002, pp. 2072–2075.

[43] L. L. Lai and J. T. Ma, "Application of evolutionary programming to reactive power planning-comparison with nonlinear programming approach," IEEE Trans. Power Syst., vol. 12, no. 1, pp. 198–206, Feb. 1997.

[44] V. Gopalakrishnan, P. Thirunavukkarasu, and R. Prasanna, "Reactive power planning using hybrid evolutionary programming method," in Proc. Power Systems Conf. & Expo. 2004 IEEE PES, New York, Oct. 10–13, 2004, vol. 3, pp. 1319–1323.

[45] J. Urdaneta, J. F. Gomez, E. Sorrentino, L. Flores, and R. Diaz, "A hybrid genetic algorithm for optimal reactive power planning based upon successive linear programming," IEEE Trans. Power Syst., vol. 14, no. 4, pp. 1292–1298, Nov. 1999.

[46] A. H. H. Al-Mohammed and I. Elamin, "Capacitor placement in distribution systems using artificial intelligent techniques," in Proc. 2003 IEEE Power Tech Conf., Bologna, Italy, Jun. 23–26, 2003, vol. 4, p. 7.

[47] Y. L. Chen and Y. L. Ke, "Multi-objective Var planning for largescale power systems using projection-based two-layer simulated annealing algorithms," IEE Proc. Generation, Transmission and Distribution, vol. 151, no. 4, pp. 555–560, Jul. 11, 2004.

[48] W. S. Jwo, C. W. Liu, C. C. Liu, and Y. T. Hsiao, "Hybrid expert system and simulated annealing approach to optimal reactive power planning," IEE Proc. Generation, Transmission and Distribution, vol. 142, no. 4, pp. 381–385, Jul. 1995.

[49] R. C. Dageneff, W. Neugebauer, and C. Saylor, "Security constrained optimization: An added dimension in utility systems optimal power flow technology," IEEE Comput. Appl. Power, pp. 26–30, Oct. 1988.

[50] W. R. Thomas, A. M. Dixon, D. T. Y. Cheng, R. M. Dunnett, G. Schaff, and J. D. Thorp, "Optimal reactive planning with security constraints," in Proc. IEEE Power Industry Computer Application Conf.,May 7–12, 1995, pp. 79–84.

[51] T. Gomez, I. J. Perez-Arriaga, J. Lumbreras, and V. M. Parra, "A security- constrained decomposition approach to optimal reactive power planning," IEEE Trans. Power Syst., vol. 6, no. 3, pp. 1069–1076, Aug. 1991.

[52] B. Cova et al., "Contingency constrained optimal reactive power flow procedures for voltage control in planning and operation," IEEE Trans. Power Syst., vol. 10, no. 2, pp. 602–608, May 1995.

[53] Torres, G.L.; Quintana, V.H.; , "An interior-point method for nonlinear optimal power flow using voltage rectangular coordinates," Power Systems, IEEE Transactions on , vol.13, no.4, pp.1211-1218, Nov 1998

[54] deOliveira, E.J.; daSilva, I.C., Jr.; Pereira, J.L.R.; Carneiro, S., Jr.; , "Transmission System Expansion Planning Using a Sigmoid Function to Handle Integer Investment Variables," Power Systems, IEEE Transactions on , vol.20, no.3, pp. 1616- 1621, Aug. 2005

# APPENDIX I

## IEEE 24 Bus System

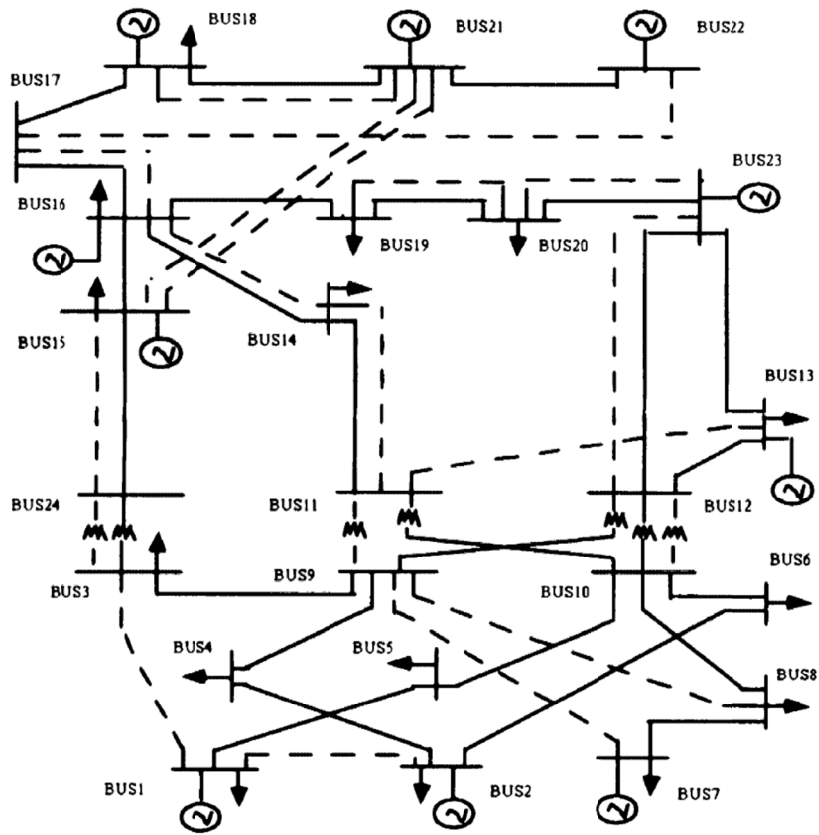

**Figure A1.1** One line Diagram of Modified IEEE 24 Bus Reliability Test System

# APPENDIX II

## IEEE 24 Bus System Specifications

**Table A.II.1 Bus Specifications of IEEE 24 bus System**

| Bus No. | Type | Vsp (pu) | Pgen + jQgen (pu) | Pdem (MW) |
|---------|------|----------|-------------------|-----------|
| 1 | Slack | 1.06 | - | 122 |
| 2 | PV | 1.04 | - | 110 |
| 3 | Load | - | - | 205 |
| 4 | Load | - | - | 85 |
| 5 | Load | - | - | 81 |
| 6 | Load | - | - | 155 |
| 7 | PV | 1.04 | - | 142 |
| 8 | Load | - | - | 195 |
| 9 | Load | - | - | 198 |
| 10 | Load | - | - | 221 |
| 11 | Load | - | - | 0 |
| 12 | Load | - | - | 0 |
| 13 | PV | 1.04 | - | 301 |
| 14 | Load | - | - | 220 |
| 15 | PV | 1.04 | - | 360 |
| 16 | PV | 1.04 | - | 115 |
| 17 | Load | - | - | 0 |
| 18 | PV | 1.04 | - | 378 |
| 19 | Load | - | - | 205 |
| 20 | Load | - | - | 145 |
| 21 | PV | 1.04 | - | 0 |
| 22 | PV | 1.04 | - | 0 |
| 23 | PV | 1.04 | - | 0 |
| 24 | Load | - | - | 0 |

**Table A.II.1** shows the bus specifications of IEEE 14 bus system. Slack and PV buses have their specified Voltages. Load Buses have their specified demands.

**Table A.II.2** shows the line parameter specifications of IEEE 14 bus system. **Table A.II.3** shows the real power generation cost characteristics.

**Table A.II.2 Line Parameter Specifications of IEEE 24 bus System**

| From Bus | To Bus | R + jX (pu) | jB/2 (pu) | Capacity (pu) |
|---|---|---|---|---|
| 1 | 2 | 0.0026+0.0139j | 0.4611j | 0.1750 |
| 1 | 3 | 0.0546+0.2112j | 0.0572j | 0.1750 |
| 1 | 5 | 0.0218+0.0845j | 0.0229j | 0.2000 |
| 2 | 4 | 0.0328+0.1267j | 0.0343j | 0.4000 |
| 2 | 6 | 0.0497+0.192j | 0.052j | 0.2000 |
| 3 | 9 | 0.0308+0.119j | 0.0322j | 0.5000 |
| 3 | 24 | 0.0023+0.0839j | 0j | 0.4000 |
| 4 | 9 | 0.0268+0.1037j | 0.0281j | 0.5000 |
| 5 | 10 | 0.0228+0.0883j | 0.0239j | 0.4000 |
| 6 | 10 | 0.0139+0.0605j | 2.459j | 0.4000 |
| 7 | 8 | 0.0159+0.0614j | 0.0166j | 0.2000 |
| 7 | 9 | 0.0159+0.0614j | 0.0166j | 0.5000 |
| 8 | 9 | 0.0427+0.1651j | 0.0447j | 0.5000 |
| 8 | 10 | 0.0427+0.1651j | 0.0447j | 0.3000 |
| 9 | 11 | 0.0023+0.0839j | 0j | 0.5000 |
| 9 | 12 | 0.0023+0.0839j | 0j | 1.0000 |
| 10 | 11 | 0.0023+0.0839j | 0j | 0.4000 |
| 10 | 12 | 0.0023+0.0839j | 0j | 0.5000 |
| 11 | 13 | 0.0061+0.0476j | 0.0999j | 0.5000 |
| 11 | 14 | 0.0054+0.0418j | 0.0879j | 0.5000 |
| 12 | 13 | 0.0061+0.0476j | 0.0999j | 0.8000 |
| 12 | 23 | 0.0124+0.0966j | 0.203j | 1.0000 |
| 13 | 23 | 0.0111+0.0865j | 0.1818j | 0.4000 |
| 14 | 16 | 0.005+0.0389j | 0.0818j | 0.4000 |
| 15 | 16 | 0.0022+0.0173j | 0.0364j | 0.9000 |
| 15 | 21 | 0.0063+0.049j | 0.103j | 1.0000 |
| 15 | 21 | 0.0063+0.049j | 0.103j | 1.0000 |
| 15 | 24 | 0.0067+0.0519j | 0.1091j | 0.5000 |
| 16 | 17 | 0.0033+0.0259j | 0.0545j | 0.5000 |
| 16 | 19 | 0.003+0.0231j | 0.0485j | 0.6500 |
| 17 | 18 | 0.0018+0.0144j | 0.0303j | 0.4000 |
| 17 | 22 | 0.0135+0.1053j | 0.2212j | 0.5000 |
| 18 | 21 | 0.0033+0.0259j | 0.0545j | 0.5000 |
| 18 | 21 | 0.0033+0.0259j | 0.0545j | 0.5000 |
| 19 | 20 | 0.0051+0.0396j | 0.0833j | 0.5000 |
| 19 | 20 | 0.0051+0.0396j | 0.0833j | 0.5000 |
| 20 | 23 | 0.0028+0.0216j | 0.0455j | 0.5000 |
| 20 | 23 | 0.0028+0.0216j | 0.0455j | 0.5000 |
| 21 | 22 | 0.0087+0.0678j | 0.1424j | 0.4000 |

**Table A.II.3 Generation Cost Characteristics of IEEE 24 bus System**

| SL No. | Type of Fuel | C(2) ($/MW$^2$ hr) | C(1) ($/MWhr) | C(0) ($/hr) |
|---|---|---|---|---|
| 1 | LNG | 86.3852 | 56.56 | 0.328412 |
| 2 | Oil | 200.6849 | 70 | 0.024142 |
| 3 | Coal | 212.3076 | 16.0011 | 0.014142 |
| 4 | Nuclear | 395.3749 | 4.4231 | 0.000213 |

# APPENDIX III

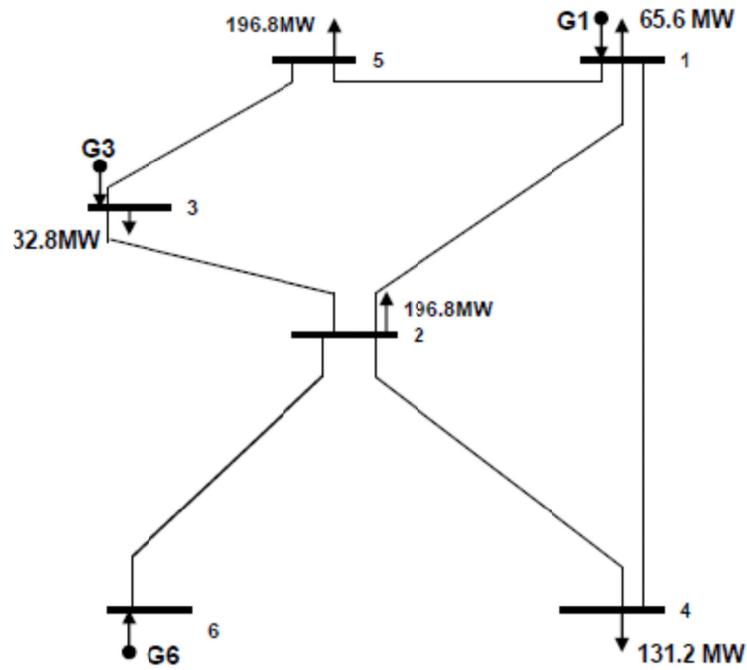

**Figure A3.1** One line Diagram of Garver's 6 bus Test System

**Table A.III.1 Bus Specifications of Garver's 6 bus test System**

| Bus No. | Type | Vsp (pu) | Pgmax | Pgmin | Qgmax | Qgmin | Pdem (MW) |
|---------|------|----------|-------|-------|-------|-------|-----------|
| 1 | Slack | 1.04 | 240 | 0 | 96 | -10 | 65.6 |
| 2 | Load | - | - | - | - | - | 196.8 |
| 3 | PV | 1.04 | 370 | 0 | 133 | -10 | 32.8 |
| 4 | Load | - | - | - | - | - | 131.2 |
| 5 | Load | - | - | - | - | - | 196.8 |
| 6 | PV | 1.04 | 610 | 0 | 196 | -10 | 0 |

**Table A.III.2 Line Parameter Specifications of Garver's 6 bus System**

| From Bus | To Bus | R + jX (pu) | jB/2 (pu) | Capacity (MVA) |
|----------|--------|-------------|-----------|----------------|
| 1 | 2 | 0.04+0.4j | 0 | 120 |
| 1 | 3 | 0.038+0.38j | 0 | 120 |
| 1 | 4 | 0.06+0.6j | 0 | 100 |
| 1 | 5 | 0.02+0.2j | 0 | 120 |
| 1 | 6 | 0.068+0.68j | 0 | 90 |
| 2 | 3 | 0.02+0.2j | 0 | 120 |
| 2 | 4 | 0.04+0.4j | 0 | 120 |
| 2 | 5 | 0.031+0.31j | 0 | 120 |
| 6 | 2 | 0.03+0.3j | 0 | 120 |
| 3 | 4 | 0.059+0.59j | 0 | 120 |
| 3 | 5 | 0.02+0.2j | 0 | 120 |
| 6 | 3 | 0.048+0.48j | 0 | 120 |
| 4 | 5 | 0.063+0.63j | 0 | 95 |
| 4 | 6 | 0.03+0.3j | 0 | 120 |
| 5 | 6 | 0.061+0.61j | 0 | 98 |

# APPENDIX IV

## Reactive Power Planning Data

Base MVA 1000 MVA

Conversion coefficient from loss to cost : 0.06$/KWH

Fixed cost associated with the installation of reactive power source : 1000$

Variable cost associated with the reactive power sources : 30$/KVAR

$q_{min}$ :Minimum size of the Capacitor  0

$q_{max}$ :Maximum size of the Capacitor  48 MVAR